\documentclass[%
 reprint,
superscriptaddress,
 amsmath,amssymb,
 aps,
]{revtex4-2}
\usepackage{comment}
\usepackage{wasysym} 

\usepackage{xr}
\usepackage{ dsfont }
\usepackage{comment}
\usepackage[normalem]{ulem}
\usepackage{soul}

\usepackage{bm}

\usepackage{graphicx}

\usepackage{color}
\definecolor{mygreen}{RGB}{98,174,37}
\definecolor{myred}{RGB}{211,0,45}
\definecolor{myblue}{RGB}{0,126,148}
\definecolor{INSBred}{RGB}{156,18,109}

\begin{document}

\preprint{APS/123-QED}

\title{
Collective deformation modes promote fibrous self-assembly in deformable particles
}


\author{Hugo Le Roy}
\email{h.leroy@epfl.ch}
\affiliation{Université Paris-Saclay, CNRS, LPTMS, 91405, Orsay, France}
\affiliation{Institute of Physics, \'Ecole Polytechnique F\'ed\'erale de Lausanne---EPFL, 1015 Lausanne, Switzerland}

\author{M. Mert Terzi}
\affiliation{Université Paris-Saclay, CNRS, LPTMS, 91405, Orsay, France}
\affiliation{PMMH, CNRS, ESPCI Paris, PSL University, Sorbonne Université, Université de Paris, F-75005, Paris, France}

\author{Martin Lenz}
\email{martin.lenz@universite-paris-saclay.fr}
\affiliation{Université Paris-Saclay, CNRS, LPTMS, 91405, Orsay, France}
\affiliation{PMMH, CNRS, ESPCI Paris, PSL University, Sorbonne Université, Université de Paris, F-75005, Paris, France}

\begin{abstract}
The self-assembly of particles into organized structures is a key feature of living organisms and a major engineering challenge. While it may proceed through the binding of perfectly matched, puzzle-pieces-like particles, many other instances involve ill-fitting particles that must deform to fit together. These include some pathological proteins, which have a known propensity to form fibrous aggregates. Despite this observation, the general relationship between the individual characteristics of the particles and the overall structure of the aggregate is not understood.
To elucidate it, we analytically and numerically study the self-assembly of two-dimensional, deformable ill-fitting particles. We find that moderately sticky particles tend to form equilibrium self-limited aggregates whose size is set by an elastic boundary layer associated with collective deformations that may extend over many particles. Particles with a soft internal deformation mode thus give rise to large aggregates. Besides, when the particles are incompressible, their aggregates tend to be anisotropic and fiber-like. Our results are preserved in a more complex particle model with randomly chosen elastic properties. This indicates that generic protein-like characteristics such as allostery and incompressibility could favor the formation of fibers in protein aggregation, and suggests design principles for artificial self-assembling structures.
\end{abstract}



\maketitle

Functional structures in living cells are often self-assembled from several copies of a single protein, from microtubules and clathrin cages to viral capsids in the shape of cylinders or spheres~\cite{Amos:1974,Pearse:1976,Caspar:1962}. The radius of such assemblies is dictated by the curvatures of the individual particles that precisely fit together to form them. Similarly, artificial self-assembly often relies on fitting well-adjusted particles together to build structures with a controlled size~\cite{Wagenbauer:2017,Sigl:2021,Hayakawa:2022,Sacanna:2010}.

In other instances however, the shapes of the individual particles are ill-fitting and do not obviously dictate the structure of the aggregate. This is the case in the pathological aggregation of normally soluble proteins, \emph{i.e.}, of proteins not evolutionarily optimized to self-assemble into a well-defined structure~\cite{Eaton:1990,Nelson:2006,Ranson:2006,Elam:2003}. Despite the diversity of the shapes and interactions involved, the aggregation of these ill-fitting proteins produces fibrous structures with remarkable consistency. These fibers display varied widths and internal structures~\cite{Bousset:2013a,Fitzpatrick:2017,Alam:2019}, and the proteins within are often significantly deformed in ways that depend on the assembly protocol~\cite{Eichner:2011}. Deformations are common in proteins, and many display physiologically relevant deformation modes that facilitate self-assembly~\cite{perottiElasticityTheoryMaturation2015,Bruinsma:2015}, perform a motor function~\cite{Chappie:2011}, participate in their biochemical activity \cite{mapaConformationalDynamicsMitochondrial2010}, or serve to mechanically transmit a signal, a function known as allostery~\cite{wodakAllosteryItsMany2019,hayer-hartlGroELGroESChaperonin2016}. Nevertheless, the generic implications of the deformability of proteins on their ill-fitting aggregation is not understood.

Beyond proteins, particle deformations have long been suggested as a mechanism to regulate aggregate size in self-assembly~\cite{Penrose:1959}. In this picture, ill-fitting particles are forced to deform as they tightly bind to one another. As more and more particles are added to the aggregate, the distortions build up until they become so severe as to prevent any further assembly~\cite{spivackStressAccumulationShape2022,Wang:2024}. The accumulation of stresses resulting from such distortions may govern the structure of DNA origami assemblies~\cite{Berengut:2020,Videbaek:2022} and prevent the indefinite bundling of preexisting protein fibers~\cite{Claessens:2008,Adamcik:2010,Brown:2014}. Beyond merely fixing the overall size of an aggregate, the accumulation of deformations can moreover dramatically alter its shape. This has been proposed to drive a transition from cylindrical to tape-like fiber bundles~\cite{Hall:2016,hallHowGeometricFrustration2017}. Finally, it can also drive sticky, deformable particles to form anisotropic aggregates that grow into infinite one-dimensional structures reminiscent of pathological protein fibers~\cite{Lenz:2017a}. The underlying mechanism and the nature of the particle properties that determine the dimensionality of the final aggregate however remain elusive.

The idea that aggregates are shaped by the frustration of their components is not limited to deformable particles, and is the object of an emergent field known as geometrically frustrated assembly~\cite{haganEquilibriumMechanismsSelflimiting2021}. Frustration can thus stem from a geometrical incompatibility between the preferred internal structure of the aggregate components, \emph{e.g.}, spheres assembling into a flat triangular lattice in 2D, and an imposed curved substrate. Such situations also lead to the formation of slender, fiber-like aggregates in theoretical models~\cite{Schneider:2005,Efrati:2013,Serafin:2021,Cheng:2023} as well as colloidal and nanoparticle experiments~\cite{Meng:2014,Yan:2020}. Systems where the frustration is carried by an additional spin-like internal degree of freedom of the particles also from slender aggregates~\cite{Hackney:2023}. Most of those designs however rely on particles with simple, regular geometrical characteristics, and little is known about the generic assembly behavior of frustrated particles with more complex properties.

In this paper, we provide a detailed analytical understanding of the emergence of self-limited and fibrous aggregates in a two-dimensional deformable particle system. We first introduce a minimal model based on highly symmetrical particles. It gives rise to an emergent elastic boundary layer length $\ell$, allowing us to map it onto a continuum description in the limit of large $\ell$. We use this description to compare the energies of several candidate structures and establish an aggregation phase diagram, which we then validate using numerical simulations. Finally, we introduce a much broader, more complex class of elastic particles, and demonstrate that the results derived in the idealized model still apply there, including in cases where the values of $\ell$ are moderate.

\section*{Elastic aggregation model}
To understand the interplay between particle deformations and aggregate structure, we first discuss a minimal one-dimensional example where ill-fitting particles deform upon aggregation. This leads to a deformation gradient from moderately deformed particles at the edge of the aggregate to highly deformed ones in its bulk. We then introduce a two-dimensional model that allows for a much wider diversity of aggregate structures, including fibers and planes. This model is analytically intractable in its general form, which leads us to design a continuum limit to enable further analysis.

\begin{figure}[t]
\includegraphics[width=8.5cm]{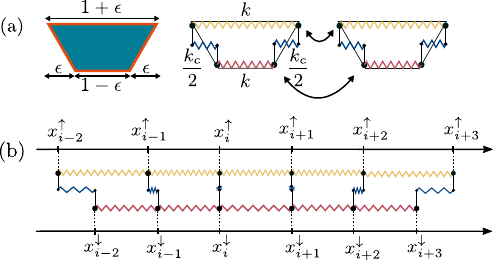}
\caption{\label{fig:1D}
The assembly of ill-fitting particles results in collective deformations, shown here in a minimal 1D model.
(a)~\emph{left:} Individual particle at rest, \emph{right:} colored springs schematizing the elasticity of the particles. In this 1D model, particles are allowed to aggregate only along the horizontal direction. Their corners then become colocalized (\emph{black arrows}), which requires deforming at least some of the springs.
(b)~Schematic of a one-dimensional particle aggregate showing the state of the springs therein. While the yellow and red springs are able to assume different lengths in the vicinity of the edge of the aggregate, they are forced to have the same lengths in the bulk. The resulting energetic penalty hampers the formation of space-filling, bulky aggregates. In two or three dimensions, a similar penalty may result in the formation of fibrous aggregates.
}
\end{figure}

\subsection*{One-dimensional toy model}
We consider a collection of identical isosceles trapezoids [Fig.~\ref{fig:1D}(a)]. Each such particle can aggregate with its left and right neighbors by fusing its vertical sides with theirs. In the special case where the trapezoids are well-adjusted, \emph{i.e.}, if they are rectangles, such binding does not require any deformation. Conversely, particles whose top and bottom faces have different lengths are ill-fitting and must deform to bind. We model the energetic cost of this deformation using four springs: two representing the top and bottom faces of the particles (yellow and red) with rest lengths $1\pm\epsilon$ and spring constant $k$, and two connecting springs with spring constants $k_c/2$ and rest lengths $\epsilon$ that tend to center the top and bottom faces (blue). Summing the contributions of these four springs, we write the deformation energy of the central particle of Fig.~\ref{fig:1D}(b) as
\begin{align}
    e_d^{(i)} = & \frac{k}{2} [(x_{i+1}^\uparrow-x_{i}^\uparrow)-(1+\epsilon)]^2
    + \frac{k}{2} [(x_{i+1}^\downarrow-x_{i}^\downarrow)-(1-\epsilon)]^2 \nonumber\\
    & + \frac{k_c}{4} (x_{i+1}^\uparrow-x_{i+1}^\downarrow-\epsilon)^2
    + \frac{k_c}{4} (x_{i}^\downarrow-x_{i}^\uparrow-\epsilon)^2,\label{eq:1D}
\end{align}
where the $\lbrace x_{i}^\uparrow\rbrace$, $\lbrace x_{i}^\downarrow\rbrace$ denote the coordinates of particle corners. This model does not involve any explicit prestresses; including some would not make any difference within the linear response regimes studied here and in the rest of this work.

Defining the shift between an upper and lower corner as $\delta_i=x_{i}^\uparrow-x_{i}^\downarrow$, force balance dictates that inside the aggregate
\begin{equation}\label{eq:force_balance}
    k ( \delta_{i+1} -2 \delta_i + \delta_{i-1}) = k_{c} \delta_i \quad \Rightarrow \quad \delta_i\propto\epsilon \sinh(i/\ell),
\end{equation}
where we define $i=0$ as the center of the aggregate and where
\begin{equation}
\ell=1/\ln\left[1+k_c/k+\sqrt{2k_c/k+(k_c/k)^2}\right]\underset{k_c\ll k}{\sim}\sqrt{{k}/{2k_c}}. 
\end{equation}
The full prefactor of the last expression of \eqref{eq:force_balance} is fixed through the force balance condition at the aggregate's left and right edges. In a large aggregate, it results in an exponential decay $\delta_i\propto\epsilon\exp(-|i-i_\text{edge}|/\ell)$ close to these edges. The initially trapezoidal particles at the center of the aggregate are thus forced into a rectangular shape ($\delta_i=0$), in contrast with the particles that reside within an edge-associated elastic boundary layer of size $\ell$.

In the limit $k_c/k\rightarrow 0$, the boundary layer size diverges. This regime is characterized by very rigid yellow and red springs, implying that the yellow and red springs close to the edge of the aggregate are almost at their equilibrium lengths. Going deeper into the aggregate, each blue spring exerts a small compressive (tensile) force on the yellow (red) chain. These forces add up over long distances, implying a progressive change of the yellow and red strain over a length scale much larger than the particle size.

\subsection*{Two-dimensional particle model}
\begin{figure}[t]
    \centering
    \includegraphics[width=9cm]{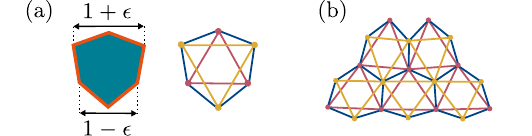}
    \caption{2D model of ill-fitting self-assembly. (a)~Individual particles have two kind of vertices, which define three interconnected elastic networks. (b)~As the particles are aggregated by matching the vertices with the same color, individual particles undergo deformations.
    }
    \label{fig:2D}
\end{figure}

One-dimensional aggregates are very simple geometrically, and are entirely characterized by the number of particles that they contain. Aggregation in higher dimension allows for a much broader variety of aggregate structures. To study the emergence of complex shapes there, we introduce a two-dimensional model based on hexagonal particles. Well-adjusted particles are represented by regular hexagons. Ill-fitting hexagons, by contrast, have alternating protruding (yellow) and withdrawn (red) corners [Fig.~\ref{fig:2D}(a)]. The corners belonging to each category form a yellow and a red equilateral triangle whose sides are springs with rest lengths $1\pm\epsilon$ and a spring constant $k$. The (blue) sides of the hexagon itself play the role of connecting springs with spring constant $k_c$, and are at rest when the yellow and red equilateral triangles are themselves at rest and centered. These hexagonal particles are three-fold symmetric, which rules out an intrinsic, particle-level preference for forming one-dimensional fibers. Any fiber formed from their aggregation will thus be an emergent symmetry-broken structure.
The response of these particles to shear and uniform compression cannot be independently varied while holding $k_c/k$ (and therefore the boundary layer size) constant. To enable particles that range from fully compressible to incompressible, we thus additionally endow both yellow and red triangles with an areal rigidity. We implement it through an energy $e^{\uparrow/\downarrow}_\text{area} = k_\text{area}(A^{\uparrow/\downarrow}-A_0^{\uparrow/\downarrow})^2/2A_0^{\uparrow/\downarrow}$, where $\uparrow$ and $\downarrow$ respectively refer to the yellow and red triangle and $A$ ($A_0$) are the associated triangle areas (rest areas).

The quadratic spring and areal energies introduced above all vanish in the particle's resting state. Any deformation away from this state implies an energetic cost, and such deformations are required to accommodate particle binding. In our model, two particles can bind along a blue side by merging one yellow and one red corner each. The merging of corners with different colors is not allowed. Each pair of bound sides is rewarded by an energy $-g$ regardless of the particles' state of deformation, which defines a zero-range interaction between particles. These rules favor the assembly of hexagons into a triangular ``particle lattice'' [Fig.~\ref{fig:2D}(b)] where all pairs of neighboring particles are bound, which we consider throughout. The aggregate topology, \emph{i.e.}, the specification of which particles bind to which others through which sides, can thus be entirely described by considering a triangular lattice and specifying a list of the lattice sites that are occupied by a particle. In the following we use the symbol $\mathcal{T}$ to refer to this topology. Since the binding energy is fully determined by the number of bound particle sides, it only depends on $\mathcal{T}$.

\subsection*{Continuum formalism}
Finding the most favorable aggregate in our 2D particle model requires two steps: to compute the optimal deformation energy for each fixed topology $\mathcal{T}$, and then to determine which topology has the lowest optimal energy. Here we introduce a continuum approximation that renders the first step analytically tractable in several important cases. This approximation is formally valid in the limit where the 2D counterpart of the boundary layer size $\ell$ is much larger than the particle size.

To define our continuum limit, we note that in large aggregates where all sites of the particle lattice discussed above are occupied (\emph{i.e.}, without holes), the yellow and red springs arrange into triangular spring lattices. In the regime $k_c\ll k$ where connecting (blue) springs are much softer than triangle (yellow and red) springs, the strain within the yellow and red triangular spring lattices varies slowly over space. This is similar to the behavior of our one-dimensional model. As a result, we can assimilate each of these triangular spring lattices to a continuum sheet, giving rise to a continuum elastic energy
\begin{align}
E_d = & \iint \left[\frac{\lambda}{2} \left(\partial_\alpha u_\alpha^\uparrow-2\epsilon\right)^2 
+ \mu\left(\frac{\partial_\alpha u_\beta^\uparrow+\partial_\beta u_\alpha^\uparrow}{2}-\epsilon\delta_{\alpha\beta}\right)^2\right]\,\text{d}A\nonumber\\
& + \iint \left[\frac{\lambda}{2} \left(\partial_\alpha u_\alpha^\downarrow+2\epsilon\right)^2
+ \mu\left(\frac{\partial_\alpha u_\beta^\downarrow+\partial_\beta u_\alpha^\downarrow}{2}+\epsilon\delta_{\alpha\beta}\right)^2\right]\,\text{d}A\nonumber\\
& + \iint\frac{\kappa_c}{2} (u_\alpha^\uparrow-u_\alpha^\downarrow)^2\,\text{d}A,\label{eq:continuum}
\end{align}
where the superscripts $^\uparrow$ and $^\downarrow$ refer to the yellow and red sheets respectively, and where the summation over repeated indices is implied while $\delta$ denotes the Kronecker delta. The displacement fields $\mathbf{u}^{\uparrow/\downarrow}(\mathbf{r})$ of either sheet are computed with respect to the infinite-aggregate, bulk state where all hexagons are regular, a state akin to a row of length-one rectangles in the one-dimensional model. Neither elastic sheet is at rest in this reference state, and $\mathbf{r}$ is the position vector in this state. The displacement gradient $\partial_\alpha u^{\uparrow/\downarrow}_\alpha$ thus plays the same role as the finite difference $(x^{\uparrow/\downarrow}_{i+1}-x^{\uparrow/\downarrow}_{i}-1)$ of \eqref{eq:1D}. The first integral of \eqref{eq:continuum} is a two-dimensional generalization of the first term of \eqref{eq:1D}, and gives the elastic energy of an isotropic elastic sheet with Lam\'e coefficients $\lambda$ and $\mu$ whose resting state is characterized by an isotropic strain $\partial_\alpha u^\uparrow_\beta=\epsilon\delta_{\alpha\beta}$. As a reminder, $\mu$ is the usual shear modulus and the sheet's Young modulus and Poisson ratio are $Y=4\mu(\lambda+\mu)/(\lambda+2\mu)$ and $\nu=\lambda/(\lambda+2\mu)$.
The integration element $\text{d}A$ runs over the reference area $A$.
Finally, the last integral captures the energy of the connecting springs. It penalize any shift between the centers of the yellow and red triangles of a given particle, and therefore any difference in the displacements of the two sheets.

The harmonic form of \eqref{eq:continuum} is valid for small deformations, which can be obtained for any aggregate topology considered here given a small enough $\epsilon$. Within this limit, the particles' binding energy can be described through
\begin{equation}\label{eq:boundary}
E_b=\gamma L,
\end{equation}
where $\gamma$ is a line tension and $L$ is the total length of aggregate edge in the reference state, including any internal holes. The parameters of the continuum model are mapped onto those of the 2D particles in the \emph{Supporting Information}, yielding
\begin{subequations}
\begin{align}
    \mu &=\frac{\sqrt{3}}{4}k\\
    \nu &= \frac{\sqrt{3}k + 2 k_{\rm area}}{3\sqrt{3}k + 2 k_{\rm area}}\\
    \kappa_c&= 2\sqrt{3}k_c\\
    \gamma&=\frac{4}{\sqrt{3}}g
\end{align}    
\end{subequations}

The binding energy of an aggregate depends only on topology $\mathcal{T}$, which we denote as $E_b(\mathcal{T})$. By contrast, the deformation energy $E_d(\mathcal{T},\lbrace\mathbf{u^{\uparrow/\downarrow}}(\mathbf{r})\rbrace)$ depends both on the topology and on the displacement fields $\mathbf{u}^{\uparrow/\downarrow}(\mathbf{r})$. As described in the beginning of this section, the optimal energy associated with a given aggregate topology is thus
\begin{equation}\label{eq:two_steps}
E(\mathcal{T}) = E_b(\mathcal{T})+\underset{\lbrace\mathbf{u}^{\uparrow/\downarrow}\rbrace}{\min}\,E_d(\mathcal{T},\lbrace\mathbf{u}^{\uparrow/\downarrow}\rbrace).
\end{equation}
Once this minimization is performed, finding the most favorable aggregate structure requires finding the topology $\mathcal{T}$ that minimizes $E(\mathcal{T})$.

\section*{Aggregation phase diagram}
To establish an aggregation phase diagram, we consider a system with a fixed but large number of particles and ask which binding topology minimizes the total energy of the system. We first use our continuum formalism to compute the energies of an infinite bulk, an elongated fiber and a disk-like aggregate, thus offering a first comparison of the stability of two-, one- and zero-dimensional structures. We then numerically compare the energies of a wider range of putative aggregate structures in our discrete particle model. Finally, we confirm the converging results of these two approaches using numerical simulations devoid of \emph{a priori} constraints on the aggregate topology.

\subsection*{Continuum phase diagram}
In 2D space-filling, infinite aggregate, all particles are forced into a regular hexagonal shape. In our model this deformation is allowed at a finite deformation cost per particle, a behavior referred to as ``shape flattening'' in the geometrically frustrated assembly literature~\cite{Grason:2016}. The continuum energy of \eqref{eq:continuum} is then minimal for $\mathbf{u}^\uparrow=\mathbf{u}^\downarrow=\mathbf{0}$, yielding an optimal energy per unit reference surface
\begin{equation}\label{eq:bulk_energy}
e_\text{bulk} =\mu\frac{1+\nu}{1-\nu}4\epsilon^2.
\end{equation}
Denoting the reference area per particle by $a$, the total energy of a set of $N$ particles thus reads $Nae_\text{bulk}$ in the large-$N$ limit. The binding energy is proportional to the perimeter of the aggregate ($E_b\propto\sqrt{N}$), and is thus negligible. 

\begin{figure*}[t]
\includegraphics[width=17.8cm]{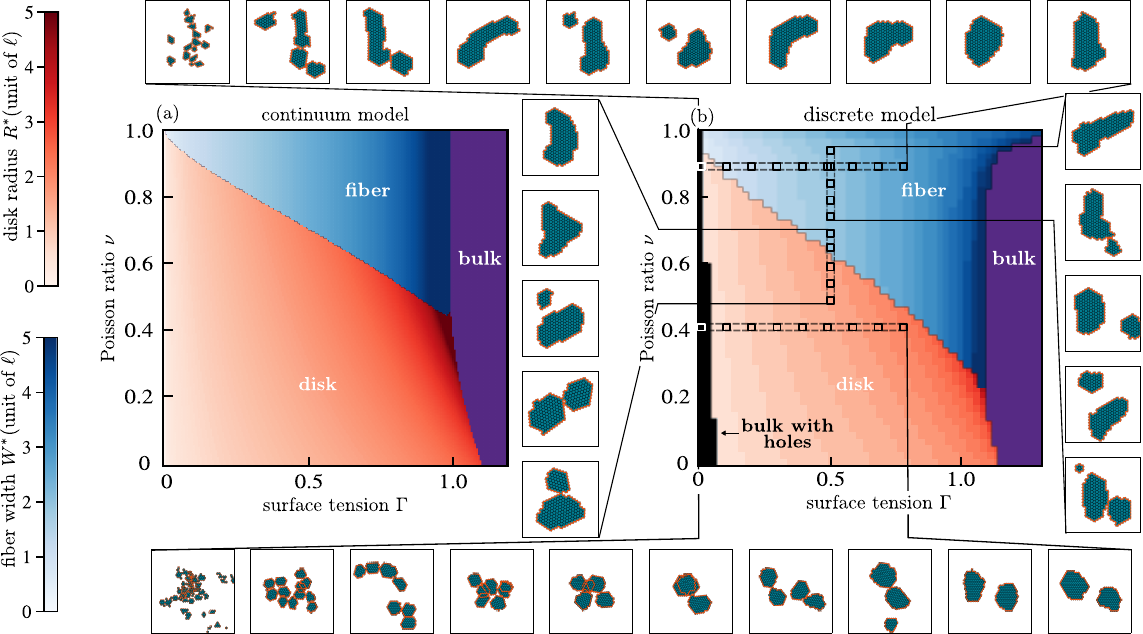}
\caption{\label{fig:phase_diagram}
Large binding energies favor bulk aggregation and incompressible particles tend to form fibrous aggregates.
(a)~Analytical phase diagram derived from Eqs.~(\ref{eq:bulk_energy}), (\ref{eq:fiber_energy}) and (\ref{eq:disk_energy}). Fiber widths always diverge when approaching the transition to the bulk. Conversely, in disks the radius discontinuously jumps from a finite value to $+\infty$ at the transition. The point where the three phases meet is $\Gamma=1$, $\nu=1/2$.
(b)~Phase diagram based on the numerical comparison of the energies of the aggregates shown in Fig.~\ref{fig:Ansatze}(a) for $\ell=5$. The fiber region of the phase diagram is larger than in the continuum model, and the disk radius $R$ again jumps discontinuously at the transition with the bulk. Smaller values of $\ell$ lead to a very similar phase diagram, albeit with an extended ``bulk with holes'' region (\emph{Supporting Information}).
Monte-Carlo simulations for the conditions indicated by the small squares are shown as small panels, and are consistent with the phase diagram. The bottom line of snapshots uses $300$ particles and a box of $30\times 30$ sites, while the others use $200$ particles and a box of $60 \times 60$ sites.
}    
\end{figure*}

To determine whether fiber formation is favored over bulk aggregation, we minimize \eqref{eq:continuum} over $\lbrace\mathbf{u}^{\uparrow/\downarrow}(\mathbf{r})\rbrace$ for an infinite strip of width $W$ and find (\emph{Supporting Information})
\begin{equation}\label{eq:fiber_displacement}
    \mathbf{u}^\uparrow(x) = -\mathbf{u}^\downarrow(x) =
    \ell \epsilon\left[ (1+\nu)\frac{\sinh(x/\ell)}{\cosh(W/2\ell)}\right]\,\hat{\mathbf{x}},
\end{equation}
where $x$ is the direction perpendicular to the fiber and where
\begin{equation}\label{eq:ell2D_definition}
\ell=\sqrt{\frac{\lambda+2\mu}{2\kappa_c}}.
\end{equation}
The quantity $\lambda+2\mu$ is known as the P-wave modulus of the sheet, and charcterizes the cost of compressing it along one axis without allowing it to deform in the perpendicular direction. Similar to \eqref{eq:1D} and Fig.~\ref{fig:1D}(b), the profile of \eqref{eq:fiber_displacement} implies bulk-like, highly deformed particles in the center of the fiber, while close to the edges the red and yellow sheets gradually relax within a boundary layer of width $\ell$. Defining the dimensionless line tension $\Gamma=2\gamma/[(1+\nu)\ell e_\text{bulk}]$, line tension cost associated with the fibers' edges reads $E_b=2\Gamma Na/W$ and the mean energy per unit surface reads
\begin{equation}\label{eq:fiber_energy}
    \frac{e_\text{fiber}(W)}{e_\text{bulk}} = 1-(1+\nu)\frac{\tanh(W/2\ell)}{W/\ell}+ (1+\nu)\frac{\Gamma}{W/\ell}.
\end{equation}
In the  $\epsilon\rightarrow 0$, small-particle-mismatch limit, all deformation energies scale as $\epsilon^2$. Thus the parameter $\Gamma$ encloses both the $\gamma$ and the $\epsilon$ dependence of all self-assembly outcomes studied in this paper. When $\Gamma<1$, $e_\text{fiber}(W)$ displays a minimum at a finite fiber width $W^*$. This optimal width diverges in the limit $\Gamma\rightarrow 1$, and the corresponding fiber is always more stable than the bulk [Fig.~\ref{fig:phase_diagram}(a)]. 
To understand this stability, consider a semi-infinite aggregate that fills half of the plane. While its energy per unit surface far from its edge is equal to $e_\text{bulk}$, the presence of the edge brings about two energetic contributions. The first is a bare line tension cost $\gamma$ per unit edge length. The second is the deformation energy gain in the boundary layer, which is of the order of $e_\text{bulk}$ per unit area. Since the width of the layer is $\ell$, the resulting energy gain per unit edge length is of the order of $\ell e_\text{bulk}$. For $\gamma \lesssim \ell e_\text{bulk}$, forming a new edge thus results in a net energy gain. At the scaling analysis level, this is equivalent to $\Gamma<1$. This argument implies that infinite bulks can lower their energy by breaking up into fibers in this regime. However, if these fibers are made so narrow that their widths become of order $\ell$ or smaller, the boundary layers associated with their two edges start to overlap. Such narrow fibers can only claim a fraction of the deformation energy reduction described above. As a result, very narrow fibers are penalized. This implies the existence of an optimal width $W^*$ that is of order $\ell$ when $\Gamma$ is of order one but smaller than one.

In an aggregate whose resting shape is a disk of radius $R$, the displacement field is given by (\emph{Supporting Information})
\begin{equation}\label{eq:disk_displacement}
    {\mathbf{u}^\uparrow(r)}
    = \ell \epsilon\frac{(1+\nu)I_1(r/\ell)}{I_0(R/\ell)+I_2(R/\ell) + \nu[I_0(R/\ell)-I_2(R/\ell)]}\,\hat{\mathbf{r}}
\end{equation}
and ${\mathbf{u}^\downarrow(r)}=-{\mathbf{u}^\uparrow(r)}$. Here $r$ denotes the radial coordinate and the $I_\alpha$s are the modified Bessel functions of the first kind. Just like Eqs.~(\ref{eq:force_balance}) and (\ref{eq:fiber_displacement}), \eqref{eq:disk_displacement} predicts an exponential decay of the boundary displacement at the edge of large aggregates. Again, the deformation cost is smaller at the aggregate edge than in its center, yielding an average energy per unit area
\begin{align}\label{eq:disk_energy}
    \frac{e_\text{disk}(R)}{e_\text{bulk}} = 1&-\frac{I_0(R/\ell)-I_2(R/\ell)}{I_0(R/\ell)+I_2(R/\ell)+\nu[I_0(R/\ell)-I_2(R/\ell)]}\nonumber\\
    &+(1+\nu)\frac{\Gamma}{R/\ell}.
\end{align}
This expression displays an optimal finite aggregate size $R^*$ at low values of $\Gamma$. As in the fiber case, this optimal disk is more stable than the bulk up to values of $\Gamma$ of order one, although the exact criterion differs due to the curved geometry of the interface [Fig.~\ref{fig:phase_diagram}(a)].

Our phase diagram indicates that fibers are more stable than disks for large Poisson ratios, \emph{i.e.}, they are favored in the aggregation of incompressible particles (characterized by $\nu=1$ in 2D). To understand this, we compare a vertical fiber and a disk at $\nu=1$. Symmetry forbids vertical (orthoradial) displacements in the fiber (disk), allowing only horizontal (radial) displacements. However, no such displacement is possible without violating the incompressibility condition of the yellow or red sheet. As a result, the sheets must remain in their resting states, implying displacements $u_x^\uparrow=-u_x^\downarrow= 2\epsilon x$ ($u_r^\uparrow=-u_r^\downarrow=2\epsilon r$) to lowest order in $\epsilon$ with respect to the fictitious bulk reference state. All the deformation cost thus comes from the connecting springs. The resulting connecting spring energy per unit area reads $8\kappa \epsilon^2x^2$ ($8\kappa \epsilon^2r^2$), proportional to the square of the distance from the center of the aggregate. This is where the difference between fibers and disks manifests itself. The edge of a fiber is just as long as its centerline, while the center of a disk is much smaller than its perimeter. As a result, a smaller proportion of connecting springs are highly extended in the fiber than in the disk, making the former energetically cheaper. Conversely, in the limit of low Poisson ratio $\nu\rightarrow 0$, the gain per unit length of the straight, fiber-like and of the curved, disk-like boundary layers become identical. Forming such boundary layers is favorable overall for $\Gamma<1$. Since the disk tends to have more boundary layer per unit area than the fiber, it is more favorable in this limit.

\subsection*{Comparison of pre-made discrete aggregate structures}
\begin{figure}[t]
    \centering
    \includegraphics{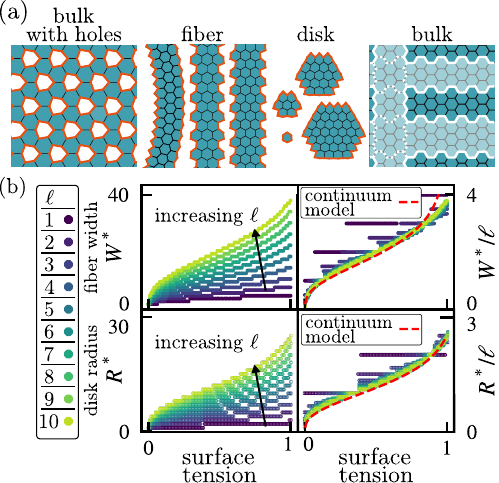}
    \caption{
    Comparison of the energies of a collection of aggregates with pre-determined topologies.
    (a)~List of the aggregate topologies included in the trial: bulk with holes, three types of fibers obtained by piling the particles in three different ways, hexagonal aggregates (including single particles) and bulk. The white lines and shading in the bulk show the three different types of cuts used to produce the fibers shown on the left. In practice, the isolated particles and uncurved fibers are never the most stable structures. We compare a broad range of fiber widths and hexagonal aggregates, despite representing only a few here.
    (b)~\emph{left:} When measured in units of number of particles, the optimal fiber width $W^*$ (computed here for $\nu=0.95$) and disk radius $R^*$ ($\nu=0.2$), strongly depend on the particles' elastic properties through $\ell$. \emph{Right:} Rescaling both lengths by $\ell$ however leads to an excellent collapse with the analytical prediction (dashed red line), even for small $\ell$.
    }
    \label{fig:Ansatze}
\end{figure}

The phase diagram of Fig.~\ref{fig:phase_diagram}(a) focuses on continuum sheets, leaving open the question of whether the formation of holes within the aggregate or regimes where the particle size is comparable to the boundary layer thickness could result in different aggregation behaviors. To assess its robustness to these effects, we numerically implement our discrete particle model in a computer. We consider several aggregates with predetermined topologies including periodic bulks with and without holes, fibers of various widths as well as hexagonal aggregates approximating disks of different radii [Fig.~\ref{fig:Ansatze}(a)]. Isolated particles are taken into account as disks with one particle.  We use a conjugate gradient algorithm to minimize the energy of each aggregate over the position of the all particle corners, analogous to the minimization over the deformation field in \eqref{eq:two_steps}. We build three different fiber structures by cutting the bulk along distinct directions. One of the fibers is left-right asymmetric and spontaneously curves, although that curvature vanishes in the large-$W$ limit. In practice we obtain the energy of infinitely long fibers by extrapolating from long ones with increasing lengths.

To compute the aggregation phase diagram, we minimize the deformation energy of aggregates with width (radii) ranging from 1 to 35 (25) for $\ell=5$ and for values of $\nu$ ranging from 0 to 1. For each value of $\Gamma$ in Fig.~\ref{fig:phase_diagram}(b) we select the aggregate with the lowest total energy. Bulks with holes, which in our model have zero deformation energy, dominate the assembly only at very small surface tensions. The rest of the phase diagram is essentially identical to the continuum one, except for an expansion of the fiber region against both bulks and disks. This may be due to the increased stability of curved fibers, which are always more stable than their straight counterparts. 

To assess the consistency of the morphology of finite-$\ell$ aggregates with the continuum ($\ell\rightarrow\infty$) expectation of Fig.~\ref{fig:phase_diagram}(a), we numerically determine the most favorable fiber width and disk radius for a range of $\ell$ and $\Gamma$ in Fig.~\ref{fig:Ansatze}(b).
While the two approaches are guaranteed to agree only in the large-$\ell$ limit, in practice the continuum approximation yields very accurate predictions all the way down to values of $\ell$ equal to the particle size. This indicates that the boundary layer physics revealed by our continuum model remains an excellent qualitative and quantitative description of the aggregation process even when the stiffness of the connecting springs $k_c$ is comparable to that of the others ($k_c\lesssim k\Leftrightarrow \ell\gtrsim 1$). The length $\ell$ thus provides a robust tool to predict the typical number of particles in the cross-section of a fiber or a disk.

\subsection*{Monte-Carlo validation of the phase diagram}
As a final validation of our phase diagram, we remove any restriction on the aggregate's structure and use a Monte-Carlo algorithm to evolve its topology. 
We simulate a triangular lattice where each site can be empty or occupied by a particle. We start with randomly placed particles, and attempt Monte-Carlo moves where a randomly chosen particle is moved to a randomly chosen empty site. We compute the optimal energy of the resulting new topology using our conjugate gradient method. The move is accepted according to a Metropolis criterion with temperature $T$. Since the deformation energy is optimized before the application of the Metropolis criterion, this temperature applies only to the system's topological degrees of freedom. To look for an approximation of the system's topological ground state, we perform a simulated annealing procedure whereby $T$ is slowly lowered from a large value to zero over the course of the simulation.

The computational cost and limited particle number in our simulation make it difficult to construct a full Monte-Carlo phase diagram. Instead, we simulate select parameter regimes to validate our main findings. Specifically, we perform three line scans at $\ell=5$ whose final aggregation states are shown in the small panels of Fig.~\ref{fig:phase_diagram}(b). First, a range of increasing $\Gamma$ at $\nu=0.9$ shows the dominance of fibers at large Poisson ratios, and the overall tendency of the fiber widths to increase with increasing line tension. A second horizontal scan at $\nu=0.4$ shows disk-like aggregates whose radii increase with increasing $\Gamma$. Finally, a vertical scan at fixed $\Gamma=0.5$ shows a transition between disks and elongated aggregates at the predicted value of $\nu$. The shapes of the aggregates resulting from the simulations are somewhat variable, as can be assessed from the $\Gamma=0.5$, $\nu=0.9$ condition which belongs to two different line scans and for which the outcomes of two independent simulations is shown in Fig.~\ref{fig:phase_diagram}(b). Our results show that despite this limitation and the relatively small number of particles in our system, our phase diagram accurately predicts the overall outcome of an unconstrained assembly.

\section*{Extension to random particles}
\begin{figure}[t]
    \centering
    \includegraphics[width=8.5cm]{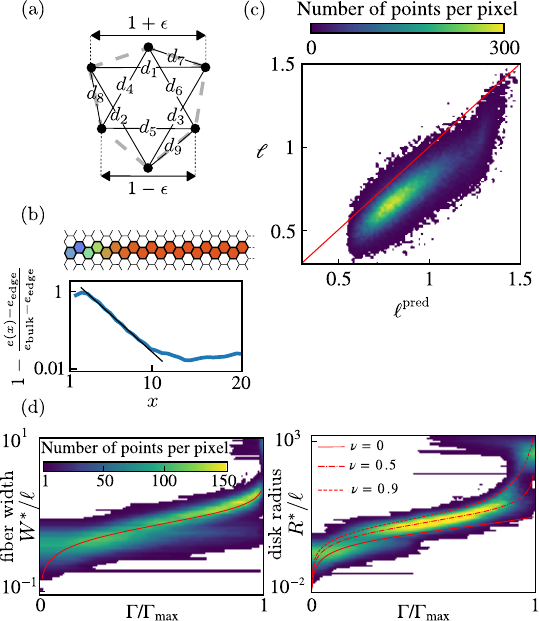}
    \caption{The aggregation behavior of particles with random elasticity is accurately captured by our continuum theory.
    (a)~The shape of a particle is fully characterized by a tuple of nine distances $(d_1,\ldots,d_9)$.
    (b)~To numerically compute the boundary layer thickness $\ell$, we simulate a half-plane full of particles (colored here by absolute deformation energy). We then perform an exponential fit of the excess elastic energy relative to the bulk as a function of distance from the edge, shown here as a line on a lin-log plot. Finally we define $\ell$ as the associated decay length.
    (c)~The random particle boundary layer thickness $\ell$ measured using this protocol is well predicted by the value $\ell^\text{pred}$ inferred from an extrapolation of \eqref{eq:ell2D_definition} (\emph{red line}).
    (d)~Radius and width of the best disk and fibers obtained for random particles. The continuum predictions are shown as red lines (the position of the line does not depend on $\nu$ for fibers).
    }
    \label{fig:random}
\end{figure}

The discrete model of Fig.~\ref{fig:2D} has been specifically designed with the continuum approximation in mind, making it unclear whether our results apply to more generic particle types. To test whether this is the case, we define a much broader class of hexagonal particles with the same shape at rest and the same binding rules but with a considerably more generic deformation energy. We parametrize the shape of each particle by nine distances $d_1$, $d_2,\ldots,d_9$ defined in Fig.~\ref{fig:random}(a). The vector $\mathbf{d}$ of these lengths completely characterizes a particle's shape, and we denote its value at rest by $\mathbf{d}_0$. The deformation energy of one particle is an arbitrary quadratic form of the deviation from that state:
\begin{equation}
    e_d=\frac{1}{2}(\mathbf{d}-\mathbf{d}_0)^T\cdot\mathbf{M}\cdot(\mathbf{d}-\mathbf{d}_0).
\end{equation}
In the following we draw the matrix $\mathbf{M}$ from a random distribution that ensures that it is positive semi-definite and three-fold symmetric (\emph{Supporting Information}). Such particles allow not only for differences in the elastic constants of the colored springs of Fig.~\ref{fig:2D}(a), but also for new couplings between them. For instance, one of these new couplings dictates that compressing a yellow spring makes the red spring to its right shrink or extend. It also opens regimes where the coupling springs are not much softer than the others.

We first determine whether the length $\ell$ in a random particle is determined by the same ratio of uniaxial-compression to triangle-shifting moduli as in \eqref{eq:ell2D_definition}. To access a wide range of $\ell$ values, we randomly draw a large number ($3\times 10^7$) of instances of matrix $\mathbf{M}$.
For each instance, we operationally define $\ell$ as the thickness of the elastic boundary layer in a simulation of a semi-infinite aggregate [Fig.~\ref{fig:random}(b)]. 
We then relate this measured value to the properties of a single particle, by numerically measuring random-particle proxies of the two moduli $\lambda+2\mu$ and $\kappa_c$  and using \eqref{eq:ell2D_definition} to compute a predicted boundary layer thickness $\ell^\text{pred}$ (\emph{Supporting Information}). As shown in Fig.~\ref{fig:random}(c), we find a good correlation between $\ell$ and $\ell^\text{pred}$ for a randomly selected subsample of $10^5$ matrices. This indicates that frustration at the edge of an aggregate relaxes in similar ways in simple and random particles, with the two subtriangles of the particles progressively shifting relative to each other. Similar to the one-dimensional model of Fig.~\ref{fig:1D}, this shift causes a restoring force (denoted $k_c\delta_i$ in the 1D model) which is balanced by stiffness of the particles' other degrees of freedom (associated with the stiffness $k$ in the 1D model). Equation~(\ref{eq:ell2D_definition}), which we use to compute $\ell^\text{pred}$, uses the P-wave modulus $\lambda+2\mu$ as a proxy for this stiffness. While this value is as an upper bound for the cost of deforming the particles, in practice our random particles may deform in more complex, less costly ways. Such deformations would result in a lower effective stiffness, and could explain why $\ell^\text{pred}$ tends to overestimate the actual value of $\ell$.

To determine whether the boundary layer thickness $\ell$ controls the aggregate size in the same way as in our simple model, we study a randomly selected subsample of $10^3$ matrices $\mathbf{M}$ from our total sample. We use the same discrete aggregate procedures as in Fig.~\ref{fig:Ansatze}(b) to determine which fiber/disk width is the most favorable for $\Gamma$ ranging from 0 to $\Gamma_\text{max}$, where we define $\Gamma_\text{max}$ as the critical line tension where the bulk becomes the most favorable structure. Both fibers and disks tend to become larger with increasing $\Gamma$ despite a large dispersion of the data similar to that observed in Fig.~\ref{fig:Ansatze}(b) (\emph{Supporting Information}). We however show in Fig.~\ref{fig:random}(d) that just as in the case of simple particles, this dispersion is largely abolished by rescaling the aggregate size by the boundary layer size $\ell$. This demonstrates that the boundary layer still controls the physics of the assembly in the random particle case. Moreover, the resulting radii and width distribution are clearly centered around the values of $W^*$ and $R^*$ predicted by the continuum limit.

\begin{figure*}[t]
    \centering
    \includegraphics[width=\textwidth]{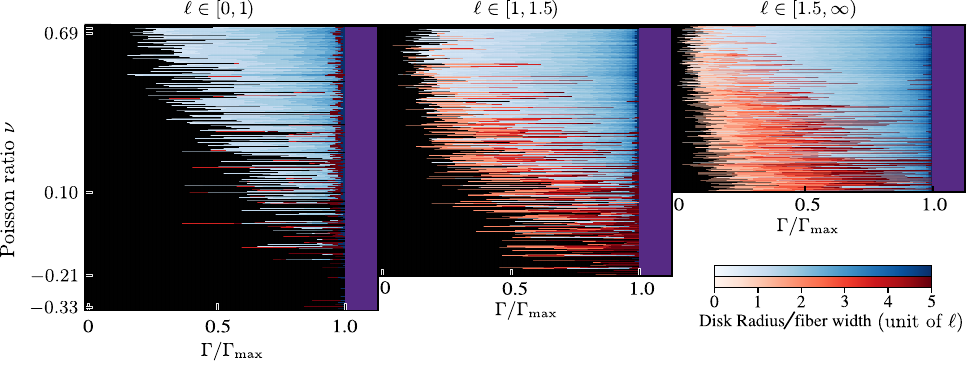}
    \caption{Random particle aggregation diagram showing good agreement with the analytical results of Fig.~\ref{fig:phase_diagram}. Color coding is as in Fig.~\ref{fig:phase_diagram}(b). Each horizontal line in the diagrams corresponds to an instance of the elasticity matrix $\mathbf{M}$. Although the boundaries between the different regions of the diagram fluctuate due to the random origin of $\mathbf{M}$, together they outline very consistent regions where bulk with holes, disks, fibers and bulks dominate.
    The diagrams with larger values of $\ell$ cover a more restricted range of $\nu$ due to the relative scarcity of particles with both large $\ell$ and small $\nu$ in matrices $\mathbf{M}$ produced by our random generation procedure. See \emph{Supporting Information} for details. 
    }
    \label{fig:random_phase_diagram}
\end{figure*}

We finally assess the applicability of the phase diagrams of Fig.~\ref{fig:phase_diagram} to random particles by correlating the most favorable aggregate type with a suitable measure of the ``individual sheet Poisson ratio'' of the analytical theory. We define this measure by decoupling the ``yellow'' and ``red'' subtriangles of our new particles from each other (\emph{i.e.}, set $M_{ij}=0$ for all $(i,j)\notin [1,3]^2 \cup [4,6]^2$), composing a lattice out of each and numerically computing their separate Poisson ratios. We use the average of these two values as our $\nu$. We then segregate our $3\times 10^7$ instances of the matrix $M$ into three groups with $\ell<1$, $1<\ell<1.5$ and $1.5<\ell$. We randomly select a few thousand particles within each group to obtain three near-uniform distribution of Poisson ratios. Fig.~\ref{fig:random_phase_diagram} shows the type and width of the aggregates obtained in each case. Fibers form at large $\nu$ and $\Gamma$ in all three groups, thus demonstrating the robust influence of these parameters. Disks also form in the expected parameter regimes, although they tend to be replaced by bulks with holes for the smallest values of $\ell$. Finally, fibers are even more predominant here than in our initial model, illustrating the broad relevance of our fiber formation mechanism upon ill-fitting self-assembly.

\section*{Discussion}
Many geometrically frustrated assembly models focus on the aggregation of ill-fitting, sticky particles with simple geometry and elasticity. Here we go one step further by investigating a simple particle geometry while allowing for arbitrary elastic properties, a small step in the direction of the complexity of proteins. Despite the added complexity, this extended particle family still displays  intelligible aggregation rules.
Compact aggregates of particles thus display a deformation gradient between a relatively unconstrained edge and a strongly deformed, frustrated core. This deformed core is energetically costly, while the aggregate's surface tension implies that its edge is also costly. As a result, the cheapest part of the aggregate lies in between the core and the edge, in the shallow bulk region. The minimization of the aggregate's energy thus requires its structure to comprise as much of this shallow bulk as possible while keeping both core and surface small. In our model, this results in aggregate size limitation and emergent anisotropy. This geometrically nontrivial optimization is \emph{a priori} strongly dependent on the elastic properties of the particles. We nevertheless identify two surprisingly simple particle-level predictors of its outcome, namely an elastic screening length $\ell$ and particle incompressibility.

In our model, the shallow bulk manifests as an elastic boundary layer with size $\ell$. This is reminiscent of the size limitation mechanisms at work in ribbons self-assembled out of planar materials with an intrinsic negative Gaussian curvature~\cite{Aggeli:2001} or specially designed warped jigsaw puzzle particles~\cite{spivackStressAccumulationShape2022}. Particles with a frustrated continuum spin-like degree of freedom and coupled incommensurate lattices in the absence of phase slips also give rise to a boundary layer~\cite{Meiri:2022,Hackney:2023,Bak:1982}. In contrast with these specific particle geometries however, here this behavior emerges in a wide variety of particles with randomly chosen elastic properties. This suggests that the mechanism may also apply in packings of complex ill-fitting proteins. Our observation of a connection between soft deformation modes at the particle level, large values of $\ell$ and consequently large self-limited aggregates thus suggests an analogy with allosteric proteins, which mechanically transmit a signal by undergoing a concerted conformational change along a soft deformation mode~\cite{thirumalaiSymmetryRigidityAllosteric2019,Bray:2004,Bray:2013}.

While extended boundary layers appear to require some particle deformation modes to be much softer than others [Fig.~\ref{fig:random}(c)], not all particle-level soft modes result in one (Fig.~S5). Indeed, once embedded in an aggregate, an initially soft deformation mode may couple to and be stiffened by the presence of neighboring particles. While our six-vertices, two-dimensional particles comprise only relatively simple soft modes compatible with the large-scale accumulation of collective deformations [namely the triangle-shifts illustrated in Fig.~S4(a-b)], more complex objects are likely to allow many more. For instance, generalizations of our model in three dimensions, where the qualitative physics highlighted here still applies, could additionally involve frustration and soft modes associated with chiral particle twisting.
Further investigations are required to identify the geometrical requirements for a soft mode to be compatible with collective deformations, which in turn allows the build-up of a thick boundary layer.
These requirements will shed light on the specific elastic parameters that most influence the particles' aggregation behavior, which could include the formation of clusters, fibers or sheets in three dimensions.
Such results could constitute a future pathway to connect our results to protein aggregation, as they would allow to assess the likelihood for a given allosteric mode to dictate the size of a frustrated protein aggregate.
This could constitute a very general tool for predicting the structure of an aggregate from the individual properties of its proteins.

The three-fold symmetry of our model particles implies that their propensity to form fibers is an emergent property as opposed to an intrinsic preference for uniaxial aggregation. This breaking of symmetry is reminiscent of the strain-induced, elongated structures formed during frustrated epitaxial growth~\cite{tersoffShapeTransitionGrowth1993,spencerMorphologicalInstabilityEpitaxially1991,eagleshamDislocationfreeStranskiKrastanowGrowth1990}. Both our most simple model and our generic, random-elasticity particles indicate that fiber formation is most advantageous in incompressible particles. This behavior is not specific to hexagonal particles, and we show in the \emph{Supporting Information} that an alternative model based on triangular objects displays an essentially identical phase diagram. 
This model design is not suitable to describe perhaps the most well-known class of protein fibers, namely those formed from amyloid proteins. Such fibers are indeed essentially stacks of beta sheets formed by unfolded sections of the protein. This implies a strong preference for piling the particles on top of one another, and thus excludes the symmetry-breaking mechanism inherent to our model~\cite{Chiti:2006,Sawaya:2007}.
By contrast, our results could be more relevant for the aggregation of dense globular proteins, which often present more than just two potential binding sites~\cite{Garcia-Seisdedos:2017}. Indeed, in some cases one or the other of these competing binding sites are favored in two closely related versions of the proteins found in different species~\cite{lynch2017human}. Such globular proteins tend to be largely incompressible and many form fibers in disease. Examples include sickle cell anemia and amyotrophic lateral sclerosis~\cite{Eaton:1990,Elam:2003,Ranson:2006}. The binding free energies involved in these assemblies are typically at least one order of magnitude larger than the thermal energy $k_BT$, justifying the analogy with our zero-temperature model.
The particle deformations in these examples moreover remain modest, consistent with the analytically tractable, small-frustration ($\epsilon\ll 1$) regime studied here.
In more asymmetric particles than those studied here, the mechanism outlined in our work could be complemented by fiber-formation mechanisms based, \emph{e.g.}, on the presence of two specific binding sites on either side of the particle. Kinetic effects such as diffusion-limited aggregation, which hampers the formation of bulky aggregates, may also favor fibers. 

Beyond proteins, the principles outlined here could be harnessed to control the assembly of artificial nano-objects. In DNA origami, soft deformation modes can be engineered to control aggregate size in a simple one-dimensional chain~\cite{Berengut:2020}. Two- or three-dimensional extensions of such designs should be prone to frustration-induced fibrillation. Fibrous morphologies also emerge in DNA origami systems into which this feature is not intentionally designed~\cite{Tikhomirov:2017}. While systems involving rigid nanoparticles are less straightforwardly mapped onto an elastic continuum than DNA origami, fibrous morphologies have been observed in packings of tetrahedral particles~\cite{Yan:2020} and successfully rationalized with an elastic model~\cite{Serafin:2021} based on a elastic frustration originating from a metric incompatibility. Our study does not rely on this very strong, somewhat specialized type of frustration and thus demonstrates that it is not required for frustration-induced fiber formation. Rigid colloids with short-range attractive and long-range repulsive interactions also display frustration-induced fibrous structures~\cite{sciortinoOneDimensionalClusterGrowth2005}. In this case, the importance of the distance dependence of the particle interaction profile falls outside of the scope of our small-frustration ($\epsilon\ll 1$) formalism. More generally, frustration build-up in the presence of nonlinear elasticity and strain-induced particle unbinding can lead to the emergence of new aggregate patterns that have only begun to be explored~\cite{Hall:2023} and could play a crucial role in the physical implementation of the principles described here.

\begin{acknowledgements}
\subsection{Funding}
ML was supported by Marie Curie Integration Grant PCIG12-GA-2012-334053, “Investissements d’Avenir” LabEx PALM (ANR-10-LABX-0039-PALM), ANR grants ANR-15-CE13-0004-03, ANR-21-CE11-0004-02, ANR-22-ERCC-0004-01 and ANR-22-CE30-0024-01, as well as ERC Starting Grant 677532. ML’s group belongs to the CNRS consortium AQV.
\subsection{Author Contributions}
M. Mert Terzi and Hugo Le Roy contributed equally to this work.
MMT performed the analytical calculations and HLR conducted the numerical simulations under the supervision of ML. All authors participated in designing the project and writing the paper.
\subsection{Competing Interests}
All authors declare no competing interests.
\subsection{Data and Materials Availability}
All the python and C++ codes used in this articles are available on github:
\begin{itemize}
\item The simulation is separated in a C++ code that minimize the elastic energy of an aggregate given a topology:
\begin{itemize}
    \item for random particles:
        \url{github.com/HugoLeRoy94/ModuleRandomHexagon.git}
    \item for hexagons:
        \url{github.com/HugoLeRoy94/Hexagone.git}
    \item for triangles:
        \url{github.com/HugoLeRoy94/Module_cpp.git}
\end{itemize}
\item The C++ codes can be used as a python object using the ctypes library. The associated python module to perform the Monte-Carlo simulation is available at:
    \url{github.com/HugoLeRoy94/Canonical.git}
\item Additionally, we designed several computations method (for the phase diagram): 
    \url{github.com/HugoLeRoy94/Extra_Module_py.git} and
    \url{github.com/HugoLeRoy94/GeneratePhaseDiagram.git}
\item Finally, the algorithms to study the random particle model are available here:
    \url{github.com/HugoLeRoy94/NU_Fiber.git}
\end{itemize}
\end{acknowledgements}
\nocite{seung1988defects}

\bibliography{Collective_deformation_modes_promote_fibrous_self-assembly_in_protein-like_particles}

\end{document}


\title{Supplementary material\\
Collective deformation modes promote fibrous self-assembly in protein-like particles}

\author{Hugo Le Roy}
\email{h.leroy@epfl.ch}
\affiliation{Université Paris-Saclay, CNRS, LPTMS, 91405, Orsay, France}
\affiliation{Institute of Physics, \'Ecole Polytechnique F\'ed\'erale de Lausanne---EPFL, 1015 Lausanne, Switzerland}

\author{M. Mert Terzi}
\affiliation{Université Paris-Saclay, CNRS, LPTMS, 91405, Orsay, France}
\affiliation{PMMH, CNRS, ESPCI Paris, PSL University, Sorbonne Université, Université de Paris, F-75005, Paris, France}

\author{Martin Lenz}
\email{martin.lenz@universite-paris-saclay.fr}
\affiliation{Université Paris-Saclay, CNRS, LPTMS, 91405, Orsay, France}
\affiliation{PMMH, CNRS, ESPCI Paris, PSL University, Sorbonne Université, Université de Paris, F-75005, Paris, France}


\maketitle
\newpage

\section{Mapping between discrete and continuous parameters}
In this section, we give the relationship between the parameters of our simple (non-random) discrete particle model and the continuum model. The former has three parameters: two spring constants $k$ and $k_c$, and an area stiffness $k_{\rm area}$. 
The continuum limit of this model comprises two coupled elastic sheets corresponding to the colors yellow and red in Fig.~2 of the main text, which we respectively denote by the $^\uparrow$ and $^\downarrow$ symbols.
We represent the elasticity of each sheet by a shear modulus $\mu$ and a Poisson ratio $\nu$. The elastic coupling between the sheets is parametrized by the coupling constant $\kappa_c$. Here we determine $\mu$, $\nu$ and $\kappa_c$ in terms of $k$, $k_c$ and $k_{\rm area}$.

We first map the energy of a single triangular spring network in the discrete particle model onto the energy of a single sheet in the continuum model. The corresponding continuum sheet energy density reads
\begin{equation}\label{eq:sheet}
    {f}_{\rm sheet}(\mu,\nu, \lbrace u_{\alpha\beta}\rbrace) = \mu \left(\frac{\nu}{1-\nu}u_{\alpha\alpha}^2+u_{\alpha\beta}u_{\alpha\beta}\right),
\end{equation}
where $u_{\alpha\beta}$ denotes the linearized strain tensor and summation over repeated indices is implied. This strain is expressed with respect to the resting configuration of the sheet of interest, and not with respect to the bulk configuration as in Eq.~(4) of the main text. The connection between the two conventions is given by the substitutions
\begin{subequations}\label{eq:sheet_strains}
    \begin{align}
        u_{\alpha\beta}&=\frac{\partial_\alpha u^\uparrow_\beta+\partial_\beta u^\uparrow_\alpha}{2}-\epsilon \delta_{\alpha\beta} \qquad\text{for the yellow sheet}\\
        u_{\alpha\beta}&=\frac{\partial_\alpha u^\downarrow_\beta+\partial_\beta u^\downarrow_\alpha}{2}+\epsilon \delta_{\alpha\beta} \qquad\text{for the red sheet.}
    \end{align}
\end{subequations}
To lowest order in $\epsilon$, the expression of the energy density of \eqref{eq:sheet} does not depend on whether it is defined as the energy per unit surface of the sheet in its own resting state or in the bulk state defined in the main text.

The energy of the discrete triangular spring network is the sum of two parts: the springs and areal energies. The springs part yields the following contribution to the sheet energy density [58].
\begin{equation}\label{eq:triangles}
    f_{\rm spring}(\lbrace u_{\alpha\beta}\rbrace) = {f}_{\rm sheet}(\mu_0,\nu_0, \lbrace u_{\alpha\beta}\rbrace)
    \qquad\text{with}\qquad
    \mu_0 = \frac{\sqrt{3}}{4}k, \qquad \nu_0 = \frac{1}{3}.
\end{equation}
Now focusing on the areal stiffness part, we find that its contribution to the total sheet energy reads
\begin{equation}
    F_{\rm area} = \frac{1}{2}k_{\rm area} \displaystyle\sum_{\rm triangles} \frac{(A - A_0)^2}{A_0}
    \approx \frac{k_{\rm area} A_0}{2} \displaystyle\sum_{\rm triangles} (u_{\alpha\alpha})^2,
    \label{eqn:energy-area}
\end{equation}
where $A_0$ is the area of a triangle and where we have used the small-deformation approximation of the relative area change as the trace of strain tensor: $(A - A_0)/A_0\approx u_{\alpha\alpha}$. The corresponding energy density in the continuum limit then reads
\begin{equation}\label{eq:areas}
f_{\rm area} = \frac{k_{\rm area}}{4} (u_{\alpha\alpha})^2,
\end{equation}
where additional factor of $1/2$ comes from the fact that only half of the triangles are endowed with an areal stiffness in our model.
Adding the two contributions of Eqs.~(\ref{eq:triangles}) and (\ref{eq:areas}), we find a total sheet energy density of the form \eqref{eq:sheet} with
\begin{equation}\label{eq:mu_nu}
\mu = \frac{\sqrt{3}}{4} k, \qquad \nu = \frac{\sqrt{3}k + 2 k_{\rm area}}{3\sqrt{3}k + 2 k_{\rm area}}.
\end{equation}

The second contribution to the total energy stems from the coupling springs. In the discrete particle model, the total coupling energy reads
\begin{equation}
    F_c = \frac{k_c}{2} \sum_\text{coupling springs} (r - r_0)^2,
\end{equation}
where $r$ and $r_0$ are the deformed and rest length of the coupling springs, respectively. To lowest order in displacement, the change of length of a spring whose direction is given by the unit vector $\hat{\mathbf{s}}$ reads
\begin{equation}
    r-r_0 \sim \left(\mathbf{u}^\uparrow - \mathbf{u}^\downarrow\right)\cdot\hat{s} = |\mathbf{u}^\uparrow - \mathbf{u}^\downarrow|\,\cos(\theta_u-\theta_s),
\end{equation}
where $\theta_u$ and $\theta_s$ are the angles that the vectors $\mathbf{u}^\uparrow - \mathbf{u}^\downarrow$ and $\hat{s}$ respectively make with the horizontal axis. As required for the continuum limit approach, we assume that the displacement fields $\mathbf{u}^\uparrow$, $\mathbf{u}^\downarrow$ are homogeneous. This yields a total energy per spring
\begin{equation}
    F_c = \frac{k_c}{2} \sum_{\text{springs }s} \left(\mathbf{u}^\uparrow - \mathbf{u}^\downarrow\right)^2 \cos(\theta_u-\theta_s)
        = \frac{N_sk_c}{2} \left(\mathbf{u}^\uparrow - \mathbf{u}^\downarrow\right)^2 \left\langle\cos^2(\theta_u-\theta_s)\right\rangle ,
\end{equation}
where $N_s$ is the total number of springs in the system and where the average $\langle\cdot\rangle$ is performed over all six possible orientations of the coupling springs, namely $\theta_s=i\pi/3$ with $i=1..6$. This averaging of the square cosine yields a result that is independent of $\theta_u$. Finally, dividing $F_c$ by the total area of the system and noting that there are six coupling springs per hexagon, we find a coupling energy per unit area
\begin{equation}
    f_c=\frac{2\sqrt{3}k_c}{2}\left(\mathbf{u}^\uparrow - \mathbf{u}^\downarrow\right)^2.
\end{equation}
Identifying this expression to the last term of Eq.~(4) of the main text, we thus find a continuum coupling constant
\begin{equation}\label{eq:kappa_c}
\kappa_c =  2 \sqrt{3} k_c .
\end{equation}

In the main text, we use Eqs.~(\ref{eq:mu_nu}) and (\ref{eq:kappa_c}) to compute values of $\nu$, $\Gamma$ and $\ell$ associated with discrete particles and compare the results of our numerical results to continuum predictions in Figs.~3 and 4.

\section{Continuum limit computations}
Here we derive the displacement fields of Eqs.~(9) and (12) of the main text from the energy functional presented in Eq.~(4) of the main text. In Sec.~\ref{sec:force_balance} we derive the general form of the force balance equations associated with the yellow and red sheets. In Sec.~\ref{sec:filament}, we solve the force balance equations in a fiber geometry. Then we solve them in a disk geometry in Sec.~\ref{sec:disk}. The expressions of the aggregate energies are computed by inserting these results into Eq.~(4) of the main text and performing the integration. The bulk elastic energy is trivially derived from either one of the resulting expressions by taking the infinite-size limit.

\subsection{Force balance equations}\label{sec:force_balance}
We differentiate the sheet energy density of \eqref{eq:sheet} with respect to the linearized strain tensor to obtain the constitutive equations of the yellow and red sheets
\begin{subequations}\label{eqn:ss-relations}
    \begin{align}
        \sigma^\uparrow_{\alpha \beta} &=
            \lambda (\partial_\gamma u^\uparrow_{\gamma}-2\epsilon) \delta_{\alpha \beta} + \mu (\partial_\alpha u^\uparrow_{\beta} +\partial_\beta u^\uparrow_{\alpha}-2\epsilon\delta_{\alpha\beta})\\
        \sigma^\downarrow_{\alpha \beta} &=
        \lambda (\partial_\gamma u^\downarrow_{\gamma}+2\epsilon) \delta_{\alpha \beta} + \mu (\partial_\alpha u^\downarrow_{\beta} +\partial_\beta u^\downarrow_{\alpha}+2\epsilon\delta_{\alpha\beta}),
    \end{align}
\end{subequations}
where $\lambda=\mu\nu/(1-\nu)$ is the first Lam\'e coefficient and where we have used the strain notation of \eqref{eq:sheet_strains}. Differentiating Eq.~(4) of the main text with respect to the yellow and red sheet displacements $u^\uparrow_{\alpha}$ and $u^\downarrow_{\alpha}$ respectively yields the force balance equations for the yellow and red sheets:
\begin{subequations}\label{eq:force_balance}
    \begin{align}
        \partial_\beta\sigma^\uparrow_{\alpha\beta}&=-\kappa_c (u^\downarrow_\alpha-u^\uparrow_\alpha)\\
        \partial_\beta\sigma^\downarrow_{\alpha\beta}&=-\kappa_c (u^\uparrow_\alpha-u^\downarrow_\alpha),
    \end{align}
\end{subequations}
whose right-hand sides represent the areal densities of external forces exerted by one sheet onto the other through the coupling springs. 

We parametrize all displacements and stresses by the bulk position vector $\mathbf{r}$. This vector is defined as the position of a point in the bulk state, \emph{i.e.}, in the state characterized by $\mathbf{u}^\uparrow=\mathbf{u}^\downarrow=0$. In other words, the actual position of any point of the yellow sheet is given by $\mathbf{r}+\mathbf{u}^\uparrow(\mathbf{r})$, and that of a point of the red sheet is given by $\mathbf{r}+\mathbf{u}^\downarrow(\mathbf{r})$. In the following, we endeavor to solve the system of equations Eqs.~(\ref{eqn:ss-relations}-\ref{eq:force_balance}) for the displacement fields $\mathbf{u}^\uparrow(\mathbf{r})$, $\mathbf{u}^\downarrow(\mathbf{r})$ on a two-dimensional domain $\Omega$ with the stress-free boundary condition
\begin{equation}\label{eq:general_BC}
    \forall \mathbf{r}\in\partial\Omega \quad n_\alpha(\mathbf{r})\sigma^\uparrow_{\alpha\beta}(\mathbf{r})=n_\alpha(\mathbf{r})\sigma^\downarrow_{\alpha\beta}(\mathbf{r})=0,
\end{equation}
where $\mathbf{n}(\mathbf{r})$ denotes the normal to the domain at a point $\mathbf{r}$ of the domain boundary $\partial\Omega$. We use linear elasticity throughout.

\subsection{Fiber}\label{sec:filament}
To study the elastic energy of an infinitely long, straight fiber, we write the position vector $\mathbf{r}=(x,y)$ in cartesian coordinates. We then solve the force balance equations over the domain $(x,y)\in \Omega= [-W/2,W/2]\times\mathds{R}$, where $W$ denotes the width of the fiber.

By translational symmetry, the displacement fields gradients $\partial_\alpha u^{\uparrow/\downarrow}_\beta$ may only depend on the horizontal coordinate $x$. this implies that the vertical displacements $u^\uparrow_y$ and $u^\downarrow_y$ are affine functions of the vertical coordinate $y$. To prevent a divergence in the coupling spring energy, the slopes of these functions must moreover be identical, and we thus write
\begin{equation}\label{eq:affine_uy}
    u_{y}^\uparrow = u_{y}^\downarrow = \phi y ,
\end{equation}
where $\phi$ is an undetermined constant that cannot depend on $x$ lest diverging strains appear in regions of large $y$ in either or both sheets. The horizontal displacements $u_{x}^\uparrow$, $u_{x}^\downarrow$ must be independent of $y$ for the same reason. Additive constants on the right-hand-side of \eqref{eq:affine_uy} can be ignored without loss of generality through suitable choices of the origins of $y$ and $\mathbf{u}^{\uparrow/\downarrow}$. As a consequence of \eqref{eq:affine_uy}, the vertical component of the force exerted by the coupling springs vanishes everywhere.

Inserting these results into Eqs.~(\ref{eqn:ss-relations}-\ref{eq:general_BC}) yields a system of coupled equations for two functions of one variables, namely $u^\uparrow_x(x)$ and $u^\downarrow_x(x)$:
\begin{subequations}\label{eq:fiber_bulk}
    \begin{align}
        \partial^2_x u^\uparrow_x &= \frac{\kappa_c}{\lambda+2\mu} (u^\uparrow_x-u^\downarrow_x)\\
        \partial^2_x u^\downarrow_x &= \frac{\kappa_c}{\lambda+2\mu} (u^\downarrow_x-u^\uparrow_x),
    \end{align}
\end{subequations}
with boundary conditions
\begin{subequations}\label{eq:fiber_boundary}
    \begin{align}
        \partial_x u^\uparrow_x (\pm W/2) &= \frac{(\lambda+\mu)2\epsilon}{\lambda+2\mu}-\frac{\lambda\phi}{\lambda+2\mu}\\
        \partial_x u^\downarrow_x (\pm W/2) &= -\frac{(\lambda+\mu)2\epsilon}{\lambda+2\mu}-\frac{\lambda\phi}{\lambda+2\mu},
    \end{align}
\end{subequations}
which is a continuum version of the one-dimensional toy model of the main text, except for the unknown constant $\phi$. 

To determine the value of $\phi$, we note that the fiber as a whole is not subjected to any external force. This implies that its total vertical tension must be constant. Due to the no-stress boundary condition at the $y=\pm\infty$ ends of the fiber, this constant is moreover equal to zero:
\begin{equation}\label{eq:longitudinal_tension}
    0= \int_{-\infty}^{\infty}\left[\sigma^\uparrow_{yy}(x)+\sigma^\downarrow_{yy}(x)\right]\,\text{d}x
     = 2(\lambda+2\mu)W\phi+\lambda U(W/2)- \lambda U(-W/2),
\end{equation}
where we have defined $U(x)=u^\uparrow_x(x)+u^\downarrow_x(x)$ and where the last equality was obtained by using \eqref{eq:force_balance} and performing the integration. We finally combine Eqs.~(\ref{eq:fiber_bulk}-\ref{eq:fiber_boundary}) to obtain a simple differential equation for $U(x)$:
\begin{equation}
\partial_x^2 U=0 \qquad \text{with} \qquad \partial_x U (\pm W/2) = -\frac{2\lambda\phi}{\lambda+2\mu},
\end{equation}
which implies $U(x)=-2\lambda\phi x/(\lambda+2\mu)$. Inserting this result into \eqref{eq:longitudinal_tension} yields $\phi=0$. 

Inserting the condition $\phi=0$ into the system Eqs.~(\ref{eq:fiber_bulk}-\ref{eq:fiber_boundary}), we find a linear system of differential equations without any unknown parameters. This system thus has a single solution, which can easily be verified to be Eq.~(9) of the main text.


\subsection{Disk}\label{sec:disk}
To study the elastic energy of a disk, we write the position vector $\mathbf{r}=(r,\theta)$ in polar coordinates. We then solve the force balance equations over the domain $(r,\theta)\in \Omega= [0,R]\times[0,2\pi)$, where $R$ denotes the radius of the disk.

The rotational invariance of the problem imposes $u_\theta^\uparrow=-u_\theta^\downarrow=0$, and implies that the radial displacement depends only on the radial coordinate. We must thus solve for two scalar functions of one variable, namely $u_r^\uparrow(r)$ and $u_r^\downarrow(r)$. Combining Eqs.~(\ref{eqn:ss-relations}-\ref{eq:general_BC}), we obtain
\begin{subequations}
    \begin{align}
        \partial_r^2u^\uparrow_r+\frac{\partial_ru^\uparrow_r}{r}-\frac{u^\uparrow_r}{r^2} &= -\frac{\kappa_c}{\lambda+2\mu}(u^\downarrow_r-u^\uparrow_r)\label{eq:up_disk}\\
        \partial_r^2u^\downarrow_r+\frac{\partial_ru^\downarrow_r}{r}-\frac{u^\downarrow_r}{r^2} &= -\frac{\kappa_c}{\lambda+2\mu}(u^\uparrow_r-u^\downarrow_r)
    \end{align}
\end{subequations}
with boundary conditions
\begin{subequations}\label{eq:disk_BC}
    \begin{align}
        \partial_ru^\uparrow_r(0)&=\partial_ru^\downarrow_r(0)=0\\
        \sigma^\uparrow_{rr}(R)&=(\lambda+2\mu)\partial_ru^\uparrow_r(R)+\lambda\frac{u^\uparrow_r(r)-\epsilon}{R}=0\\
        \sigma^\downarrow_{rr}(R)&=(\lambda+2\mu)\partial_ru^\downarrow_r(R)+\lambda\frac{u^\downarrow_r(r)+\epsilon}{R}=0.
    \end{align}
\end{subequations}

By summing these equations two by two, we find a system of equations for $U(r)=u^\uparrow_r(r)+u^\downarrow_r(r)$, namely
\begin{equation}
    \partial_r^2U+\frac{U}{r}-\frac{U}{r^2} = 0 \quad\text{with}\quad
    \partial_rU(0)=0 \quad\text{and}\quad
    (\lambda+2\mu)\partial_rU(R)+\lambda\frac{U(r)}{R}=0.
\end{equation}
This implies $U(r)=0$ and therefore $u^\uparrow_r(r) =-u^\downarrow_r(r)$. Plugging this condition into \eqref{eq:up_disk} yields
\begin{equation}
    \partial_r^2u^\uparrow_r+\frac{\partial_ru^\uparrow_r}{r}-\frac{u^\uparrow_r}{r^2} = \frac{2\kappa_c}{\lambda+2\mu}u^\uparrow_r,
\end{equation}
which alongside the boundary conditions on $u^\uparrow_r$ comprised in \eqref{eq:disk_BC} form a fully specified second-order linear differential equation, for which Eq.~(12) of the main text is the unique solution.

\section{Smaller $\ell$ phase diagram}
\begin{figure}[t]
    \centering
    \includegraphics{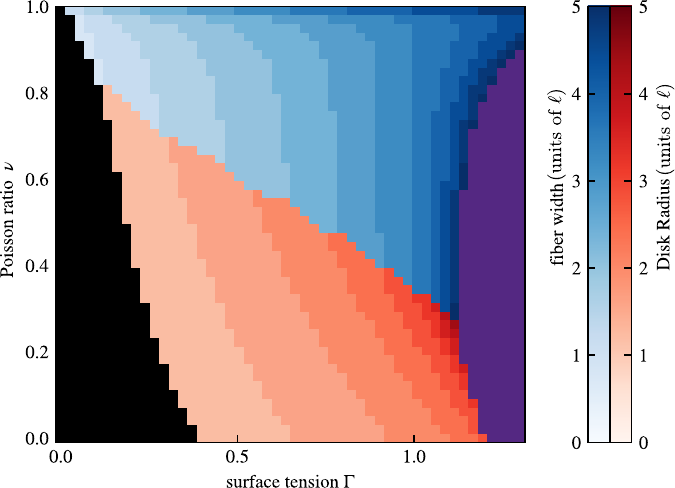}
    \caption{Phase diagram based on the numerical comparison of the energies of the aggregates shown in Fig.~4(a) of the main text for $\ell = 2.5$. The color code is as in Fig.~3 of the main text; black: bulk with holes, red: disk, blue: fiber, purple: bulk.}
    \label{fig:low_ell_phase_diagram}
\end{figure}

Figure~3 of the main text presents the phase diagram of our non-random discrete particles for boundary layer thicknesses $\ell=+\infty$ [Fig.~3(a), continuum theory] and $\ell=5$ [Fig.~3(b), numerical procedure described in the main text]. To further illustrate the influence of the boundary layer thickness $\ell$, in Fig.~\ref{fig:low_ell_phase_diagram} we follow the same procedure used to compute the diagram of Fig.~3(b) for a smaller value $\ell=2.5$. Consistent with the results obtained for random particles in Fig.~6 of the main text, we observe that smaller values of $\ell$ induce a loss of disks to the benefit of bulks with holes, but no dramatic changes in the fiber and bulk regions of the diagram.

\section{\label{sec:random_procedure}Procedure to draw random matrix}
Here we describe the procedure we use to generate the random instances of the matrix $\mathbf{M}$ introduced in Eq.~(14) of the main text.
The form of the energy chose in Eq.~(14) is very generic, as it boils down to a small-displacement Taylor expansion of any energy function of the 12 vertex coordinates under the constraints of translational and rotational invariance. Here we discuss the way in which we enforce the additional symmetries of the matrix $\mathbf{M}$.

As discussed at the end of the main text, we demand that the elasticity of our particles be three-fold symmetric, \emph{i.e.}, invariant under the permutation:
\begin{align*}
d_1 & \rightarrow d_2 \nonumber\\
d_2 & \rightarrow d_3 \nonumber\\
d_3 & \rightarrow d_1 \nonumber\\
d_4 & \rightarrow d_5 \nonumber\\
d_5 & \rightarrow d_6 \nonumber\\
d_6 & \rightarrow d_4 \nonumber\\
d_7 & \rightarrow d_8 \nonumber\\
d_8 & \rightarrow d_9 \nonumber\\
d_9 & \rightarrow d_7, \nonumber\\
\end{align*}
where the distances $d_i$ are defined in Fig.~5(a) of the main text and the permutation above applies simultaneously to the vectors $\mathbf{d}$ and $\mathbf{d}_0$ of actual and resting positions. To enforce this condition, we first draw all entries of a $9\times 9$ matrix $\mathbf{M}_0$ as independent identically distributed variables from the normal distribution $\mathcal{N}(0,1)$. We then define the $9\times 9$ block matrix $\mathbf{\Omega}$ that enforces the aforementioned permutation as:
\begin{equation}
\mathbf{\Omega} = 
\left(\begin{matrix}
  \mathbf{\omega} & 0 & 0\\
  0 & \mathbf{\omega} & 0 \\
  0 & 0 & \mathbf{\omega}
\end{matrix}\right),
\qquad\text{where the block $\mathbf{\omega}$ is given by}\qquad
\mathbf{\omega}=
\left(\begin{matrix}
  0 & 0 & 1\\
  1 & 0 & 0 \\
  0 & 1 & 0
\end{matrix}\right).
\end{equation}
We then apply permutations to the symmetry-less matrix $\mathbf{M}_0$ to obtain
\begin{equation}\label{eq:sym1}
\mathbf{M}_1 = \frac{1}{3}\left(\mathbf{M}_0 + \mathbf{\Omega} \mathbf{M}_0 \mathbf{\Omega}^{-1} + \mathbf{\Omega}^2 \mathbf{M}_0  \mathbf{\Omega}^{-2} \right),
\end{equation}
which has the required three-fold symmetry property.

Our second and last requirement for our elasticity matrix is that it be semi-positive, which prevents the ground state of our individual particles from being mechanically unstable. We enforce this condition by defining
\begin{equation}\label{eq:sym2}
\mathbf{M}=\mathbf{M}_1\mathbf{M}_1^T,
\end{equation}
where $^T$ denotes the usual matrix transposition. By combining Eqs.~[\ref{eq:sym1}] and [\ref{eq:sym2}] and realizing that $\mathbf{\Omega}^3=\mathbf{I}$ it is easy to show that the energy of Eq.~(14) of the main text then satisfies the three-fold symmetry condition $e_d(\mathbf{\Omega}\mathbf{d})=e_d(\mathbf{d})$. All elasticity matrices used in Figs.~5 and 6 of the main text are obtained through the procedure described here.

\section{Self-limited aggregate sizes scale like the boundary layer thickness $\ell$ in random particles}
\begin{figure}[t]
    \centering
    \includegraphics{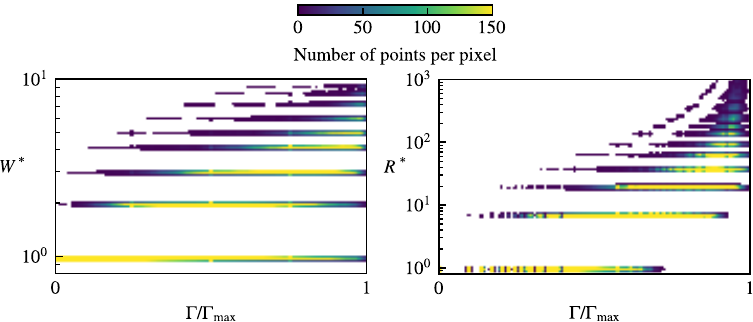}
    \caption{\label{fig:R_W_Gamma_no_rescaling}
    Equilibrium width $W^*$ and radius $R^*$ of random particles aggregates in units of number of particles. The data shown is identical to that of Fig.~5(d) of the main text, only without rescaling by the boundary layer thickness $\ell$.}
\end{figure}

Randomly drawn matrices $\mathbf{M}$ display heterogeneous elastic properties, which result in a wide distribution of aggregate sizes. In Fig.~\ref{fig:R_W_Gamma_no_rescaling} we show that the value of the surface tension $\Gamma/\Gamma_\text{max}$ is not sufficient to accurately predict the size of the aggregates resulting from a matrix $\mathbf{M}$. By contrast, we show in Fig.~5(d) of the main text that rescaling the aggregate sizes by the boundary layer thickness $\ell$ leads to a collapse of the aggregate sizes, consistent with our continuum theory.

\section{\label{sec:random_diagram_procedure}Selection of the random particles used in the phase diagrams}
\begin{figure}[b]
    \centering
    \includegraphics[width=17cm]{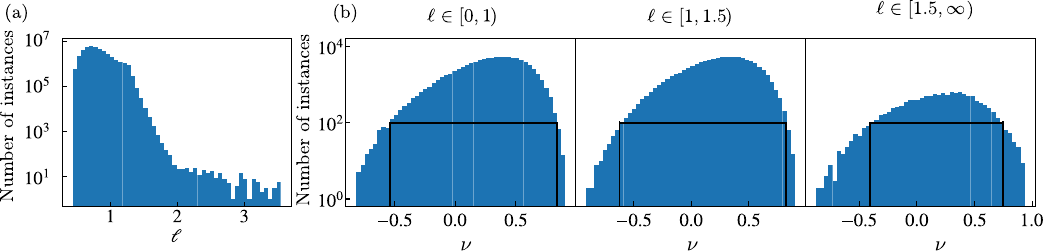}
    \caption{
    Distribution of random particle properties resulting from the procedure of Sec.~\ref{sec:random_procedure}.
    (a)~Distribution of boundary layer lengths within the initial draw of $3\times 10^7$ elasticity matrices $\mathbf{M}$.
    (b)~Distribution of Poisson ratio within each of the three batches of particles. We respectively pick 3400, 3500 and 2400 particles out of the three batches to obtain the quasi-uniform distributions of Poisson ratios materialized by the black lines.
    }
    \label{fig:histograms}
\end{figure}

To generate the random particle aggregation diagrams of Fig.~6 of the main text, we first draw $3\times10^7$ random matrices and compute the boundary layer length of the associated particles. This large sample size is required to obtain a sufficient number of particles with relatively high values of $\ell$, as is apparent from the fast decay of the probability density of Fig.~\ref{fig:histograms}(a) as $\ell$ increases.

We next divide the range of accessible boundary layer lengths into three intervals: $\ell \in [0,1)$, $\ell \in [1,1.5)$ and $\ell \in [1.5,\infty)$. We randomly select batches of $10^6$ particles from the first two intervals and use the whole third batch, which contains only $1.25 \times 10^5$ particles. We then compute the Poisson ratio (see Sec.~\ref{sec:l_pred} for the procedure) for all particles in the three batches and further select a few thousand particles from each to obtain quasi-uniform distributions of Poisson ratios as represented in Fig.~\ref{fig:histograms}(b). Finally, for each particle, we construct an aggregation diagram by numerically determining the best aggregate upon varying the surface tension. The outcome of this procedure is Fig.~6 of the main text.

\section{\label{sec:l_pred}Predicting the boundary layer thickness from the elastic properties of random particles}
\begin{figure}[t]
    \centering
    \includegraphics[width=13cm]{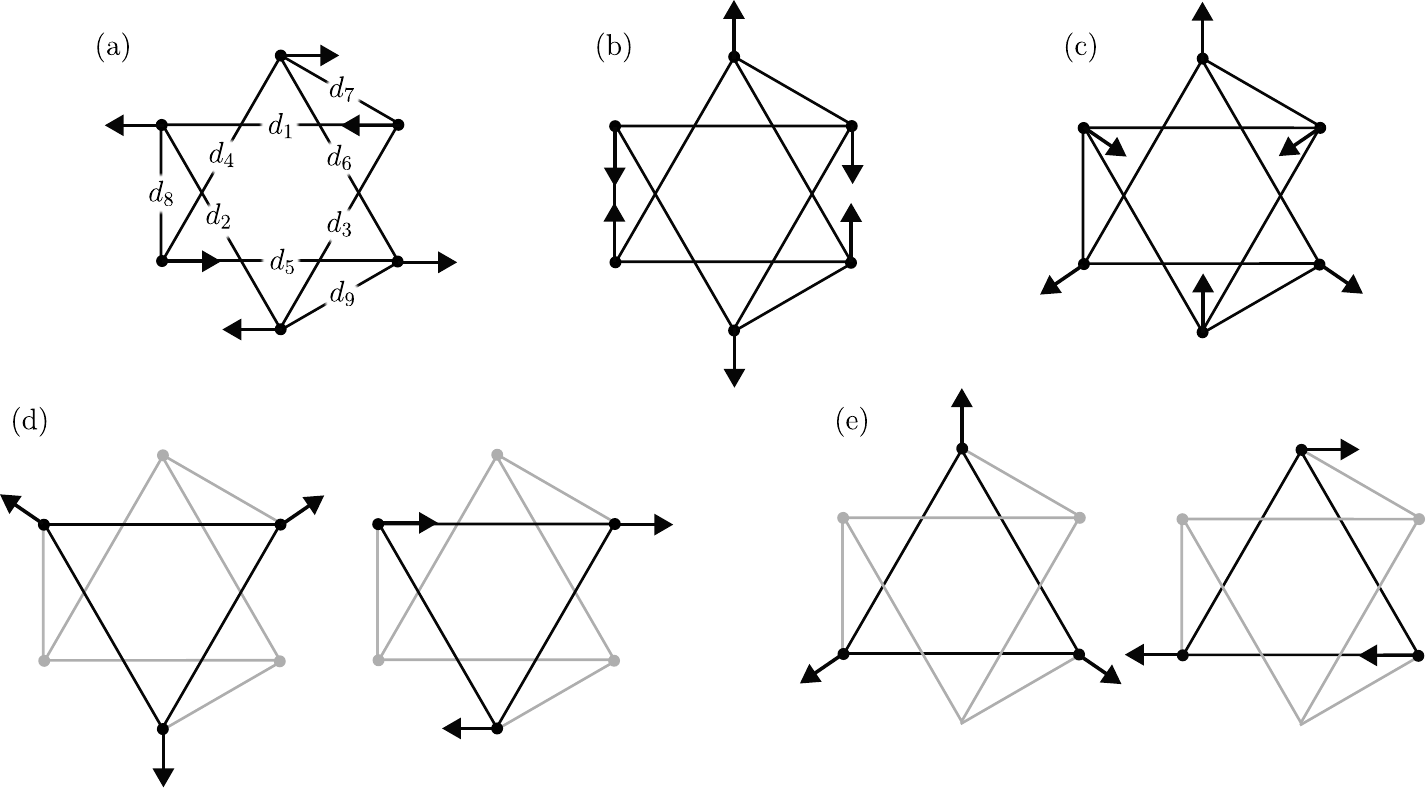}
    \caption{Deformation protocols used to estimate the boundary layer thickness and Poisson modulus of a collection of random particles. These estimates are made to zeroth order in $\epsilon$. We thus set $\epsilon$ to zero in this figure and throughout the procedure. The black arrows denote externally imposed node displacements.
    (a)~Displacement protocol used to generate the estimate $\kappa^\text{pred}_c$ of the coupling modulus in \eqref{eq:kappac_pred}. The definitions of the elements of the distance vector $\mathbf{d}$ are recalled here for convenience.
    (b)~Another displacement protocol equivalent to that of the previous panel. 
    (c)~Simultaneous isotropic bulk deformation of the two triangles used to estimate the sheet bulk modulus $K^\text{pred}$. In the case of the simple particles of Fig.~2 of the main text, this mixture of expansion and compression imposes bulk deformations on both sub-triangles while leaving the coupling springs lengths unchanged to lowest order in deformation. We thus use it to extract the value of the bulk modulus independently from the coupling modulus in \eqref{eq:K_pred}.
    (d)~Bulk and shear deformation of the yellow triangle used to compute $\nu^\uparrow$ through $K^\uparrow$ and $\mu^\uparrow$ in \eqref{eq:sheet_nus}. The grey nodes are assumed to be free to move in such a fashion that the length of the grey segments remains unchanged during the deformation. This is equivalent to decoupling the black triangle from the rest of the particle, as mentioned in the main text.
    (e)~Bulk and shear deformation of the red triangle used to compute $\nu^\downarrow$ through $K^\downarrow$ and $\mu^\downarrow$ in \eqref{eq:sheet_nus}. The meaning of the grey nodes is the same as in the previous panel.\\
    \label{fig:soft_deformation}}
\end{figure}

In our continuum model, the thickness of the boundary layer that marks the transition between strongly constrained bulk particles and relatively unconstrained aggregate-edge particles is directly tied to the ease with which the two subtriangles that constitute the particles can be shifted relative to each other. Here we tentatively apply a similar reasoning to random particles, and construct the ultimately successful estimate $\ell^\text{pred}$ of the boundary layer thickness presented in Fig.~5(c) of the main text. In our continuum model the boundary layer thickness $\ell$ is constructed from a ratio of elastic moduli, namely
\begin{equation}\label{eq:moduli_ratio}
    \ell^2=\frac{\lambda+2\mu}{2\kappa_c} = \frac{K}{(1+\nu)\kappa_c}.
\end{equation}
where $\kappa_c$ denotes the inter-sheet coupling constant, $K=\lambda+\mu$ is the intra-sheet bulk modulus and $\nu=\lambda/(\lambda+2\mu)$ is the intra-sheet Poisson ratio. These parameters are not rigorously well-defined in a bulk aggregate of random particles characterized by an elasticity matrix $\mathbf{M}$ [Eq.~(14) of the main text], which does not in general exactly map onto the continuum energy of Eq.~(4) of the main text. To nonetheless derive our estimate $\ell^\text{pred}$, here we set out to compute proxies for each of these three parameters.  We base our procedure on the computation of pseudo-moduli associated with specific deformations illustrated in Fig.~\ref{fig:soft_deformation}. We thus define the pseudo-modulus $\mathcal{K}$ associated with a deformation vector $\delta\mathbf{d}=\mathbf{d}-\mathbf{d}_0$ as
\begin{equation}\label{eq:pseudomodulus}
    \mathcal{K}[\delta\mathbf{d}] = \frac{\delta\mathbf{d}^{T}\cdot\mathbf{M}\cdot\delta\mathbf{d}}{A_0^{\rm hex}},
\end{equation}
where $A_0^{\rm hex} = \sqrt{3}/2$ is the resting area of a hexagonal particle to zeroth order in $\epsilon$. In the following, we choose the normalization of $\delta\mathbf{d}$ so that the definition of \eqref{eq:pseudomodulus} coincides with the moduli discussed in Eq.~(4) of the main text when applied to our simple particle model (Fig.~2 of the main text). In that specific case, the deformations $\delta\mathbf{d}$ used below are eigenvectors of $\mathbf{M}$.

We first estimate the sheet-coupling modulus $\kappa_c$ as the pseudo-modulus associated with a relative displacement between the two subtriangles of a particle. We picture two possible protocols for such a displacement in Fig.~\ref{fig:soft_deformation}(a) and (b), namely a horizontal or a vertical shift of the two triangles, although any intermediate direction is also allowed. Due to the three-fold symmetry of the particle, all these shifting protocols are associated with the same modulus. Using the deformation mode of Fig.~\ref{fig:soft_deformation}(a), we write
\begin{equation}\label{eq:kappac_pred}
    \kappa_c^{\rm pred} = \mathcal{K}\left[\left(0, 0, 0, 0, 0, 0, -{\sqrt{3}}/{2}, 0 , {\sqrt{3}}/{2}\right)\right].
\end{equation}

We next estimate the intra-sheet bulk modulus $K$, which characterises the stiffness of each sheet with respect to an isotropic expansion or compression, as the pseudo-modulus associated with the expansion-compression deformation illustrated in Fig.~\ref{fig:soft_deformation}(c), namely
\begin{equation}\label{eq:K_pred}
    K^\text{pred} = \mathcal{K}\left[\left(-2^{-3/2}, -2^{-3/2}, -2^{-3/2}, 2^{-3/2}, 2^{-3/2}, 2^{-3/2}, 0, 0, 0\right)\right].
\end{equation}

Our third step is to estimate the Poisson ratio as the average
\begin{equation}
    \nu^{\rm pred} = \frac{\nu^\uparrow + \nu^\downarrow}{2}
    \label{eqn:nu_estimate}
\end{equation}
of the individual pseudo-Poisson ratios $\nu^\uparrow$ and $\nu^\downarrow$ of the yellow and red sublattices (and thus of the yellow and red sheets in the continuum limit). We estimate each of these sheet-specific Poisson ratios as
\begin{equation}\label{eq:sheet_nus}
    \nu^{\uparrow/\downarrow}=\frac{K^{\uparrow/\downarrow}-\mu^{\uparrow/\downarrow}}{K^{\uparrow/\downarrow}+\mu^{\uparrow/\downarrow}},
\end{equation}
where we define the sheet-specific bulk and shear pseudo-moduli through the four deformations illustrated in Fig.~\ref{fig:soft_deformation}(d-e), namely
\begin{subequations}
    \begin{align}
        K^\uparrow     & = \mathcal{K}\left[\left(1/2, 1/2, 1/2, 0, 0, 0, 0, 0, 0\right)\right]\\
        \mu^\uparrow   & = \mathcal{K}\left[\left(0, -\sqrt{3}/4, \sqrt{3}/4, 0, 0, 0, 0, 0, 0\right)\right]\\
        K^\downarrow   & = \mathcal{K}\left[\left(0, 0, 0, 1/2, 1/2, 1/2, 0, 0, 0\right)\right]\\
        \mu^\downarrow & = \mathcal{K}\left[\left(0, 0, 0, \sqrt{3}/4,0, -\sqrt{3}/4, 0, 0, 0\right)\right].
    \end{align}
\end{subequations}

We finally combine Eqs.~(\ref{eq:kappac_pred}-\ref{eqn:nu_estimate}) by inserting them into \eqref{eq:moduli_ratio}, which yields the values of $\ell^\text{pred}$ displayed in Fig.~4(c) of the main text. We also use the pseudo-Poisson ratio defined in \eqref{eqn:nu_estimate} as the vertical coordinate of the phase diagrams of Fig.~6 of the main text.

\section{Unsuccessful alternative predictor of the boundary layer thickness in random particles}
The successful method described in Sec.~\ref{sec:l_pred} is only one of many possible extensions of our continuum theory to random particles. Here we discuss a different, unsuccessful approach and draw conclusions from its failure.

The foundation of our continuum approach is the existence of an emergent length scale in our deterministic 2D particle model that diverges in the limit where one elastic constant of the particles ($k_c$) becomes much smaller than another ($k$).  
Mechanistically, this large mismatch in elastic constants means that the restoring forces from the softer deformation mode of the particle \emph{slowly} accumulate from the edge of the aggregate to the bulk as discussed in the main text for the one-dimensional model of Fig.~1.
Mathematically, any deformation of the particle may be decomposed into a linear combination of the eigenvectors of the matrix $\mathbf{M}$ defined in Eq.~(14) of the main text. These deformation eigenmodes do not couple to each other in an isolated particle, and each has its own stiffness associated with the corresponding eigenvalue of $\mathbf{M}$. In our deterministic 2D particles, the length $\ell$ is inversely proportional to the square root of the smallest of these eigenvalues.

\begin{figure}[t]
\includegraphics[width=18cm]{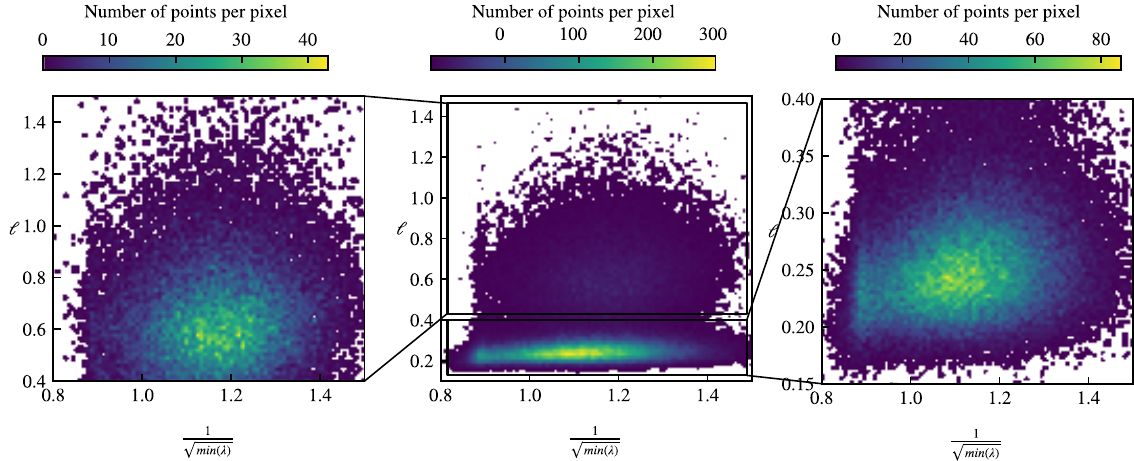}
\caption{The predictor $\ell^\text{eigen}$ of \eqref{eq:l_eigen} does not correlate with the measured boundary layer thickness $\ell$. Each of the two side panels is a close-up of a region of the central panel, as indicated by the black rectangles.
}
\label{fig:softest_mode}
\end{figure}

This leads us to hypothesize that any soft mode of deformation of the particle may be able to play the same role that the shifting mode of Fig.~\ref{fig:soft_deformation}(a-b) plays in our deterministic 2D model: to produce slowly accumulating stresses that take the particle from a more relaxed edge configuration to the bulk configuration. If several soft modes exist within the particle, the softer one should correspond to the thickest boundary layer and thus should dominate the decay of the elastic deformation far enough from the aggregate edge. This reasoning thus suggests the following proxy for the boundary layer thickness:
\begin{equation}\label{eq:l_eigen}
    \ell^\text{eigen}=\frac{1}{\sqrt{\underset{i}{\min}\,\lambda_i}},
\end{equation}
where the $\lambda_i$ denote the eigenvalues of $\mathbf{M}$. We test the accuracy of this predictor in Fig.~\ref{fig:softest_mode} using the same plotting convention as in Fig.~5(c) of the main text, and find that it does not significantly correlate with $\ell$. As detailed in the discussion section of the main text, we conclude that not all soft modes of our particles are compatible with the mechanism of stress accumulation over large length scales described above.

\section{Triangular particles}
\begin{figure}[t]
    \centering
    \includegraphics{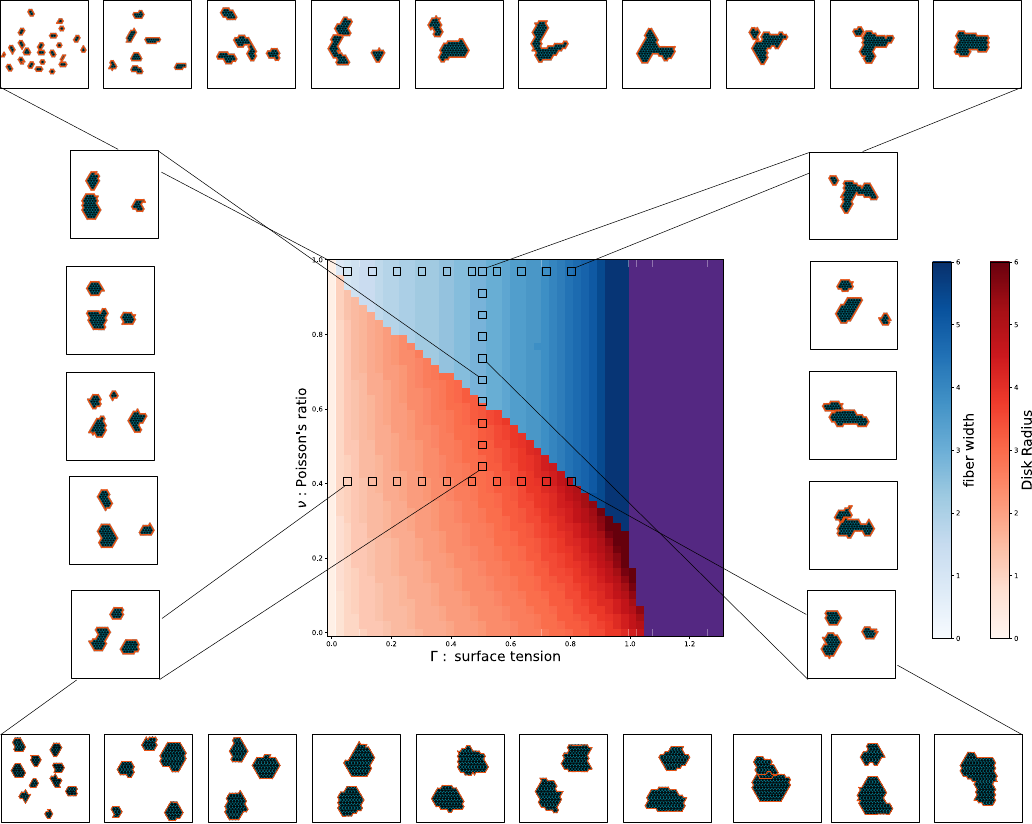}
    \caption{\label{fig:triangular_particles}Alternative triangle-based model for frustrated self-assembly.
    (a)~Particle design as in Fig.~2(a) of the main text.
    (b)~Example of mechanically equilibrated aggregates as in Fig.~2(b) of the main text. To improve visibility, coupling (blue) springs are not represented.
    (c)~In an infinite (bulk) aggregate, the length of the edge of the two triangles must match.
    (d)~Phase diagram and Monte-Carlo simulation results for $\ell=2.5$ as on the right-hand-side of Fig.~3 of the main text. The bottom line of snapshots contain $300$ particles in a box of $30\times 30$ lattice sites. The other snapshots contain $200$ particles in boxes of size $40\times 40$.
    (e)~Examples of aggregate topologies whose energies we compare to compute the phase diagram.
}
\end{figure}

To demonstrate the robustness of the results of the main text to a change in particle design, here we introduce a model of triangular deterministic particles. Similar to the main text, the model particles comprise two triangles made of hard $k$ springs and a set of softer (six in this case) $k_c$ coupling springs. The particle design is illustrated in Fig.~\ref{fig:triangular_particles}(a), and aggregates thereof are shown in Fig.~\ref{fig:triangular_particles}(b) and (c). This design gives rise to the same continuum theory as the model of the main text, and we conduct comparisons of pre-made discrete aggregate structures as well as Monte-Carlo simulations using the same methodology as in the main text. These results are shown in Fig.~\ref{fig:triangular_particles}(d). The aggregate designs used for the former type of analysis are shown in Fig.~\ref{fig:triangular_particles}(e). By contrast with the model described in the main text, the bulk with hole is never advantageous because sticking two triangular particles together has a non-zero elastic energy cost. As a result, a gas of isolated single particles is favored at very low tensions. In addition, unlike in Fig.~3(b) of the main text, the frontier between fiber and bulk is perfectly vertical. This further confirms that the extended region of stability of the fibers observed for the hexagonal model of the main text is due to the curvature of the fibers. Apart from these nuances, the results obtained with this model are very similar to those of the deterministic 2D hexagonal particle model detailed in the main text, thus confirming the robustness of our continuum approach.

\begin{table}
\centering

\begin{tabular}{l l l}
 \\
effect\_of\_ell\_2.pdf & PDF & \href{https://arxiv.org/submit/5518589/delete_file?file=effect_of_ell_2.pdf}{Delete} \\

\end{tabular}

\end{table}
\bibliography{bib-SI}